\documentclass[a4paper,11pt]{article}
\pdfoutput=1 

\usepackage{jcappub} 

\usepackage[T1]{fontenc} 

\title{\boldmath Gravitational Lensing of Schwarzschild and Charged Black Holes Immersed in Perfect Fluid Dark Matter Halo}

\author[a]{Chen-Kai Qiao,}
\author[a]{Mi Zhou,}

\affiliation[a]{College of Science, Chongqing University of Technology, Banan, Chongqing, 400054, China}

\emailAdd{chenkaiqiao@cqut.edu.cn}
\emailAdd{lilyzm@cqut.edu.cn}

\abstract{Dark matter and dark energy dominate the behavior of our universe. The dark matter usually forms halo structures in large number of galaxies. Properties of dark matter halo can be revealed and understood from the gravitational lensing observations. In this work, a comprehensive study on the gravitational lensing of black holes immersed in dark matter halos is presented. To effectively model the supermassive black hole in a galaxy center (which is surrounded by dark matter halo) in a simple way, we investigate the Schwarzschild black hole and charged Reissner-Nordstr\"om black hole immersed in a perfect fluid dark matter halo. In the present work, several basic quantities in gravitational lensing (the gravitational deflection angle of light, photon sphere, black hole shadow radius, gravitational lens equation and Einstein ring) are calculated and analyzed analytically and numerically. A second order analytical expansion of gravitational deflection angle is obtained in the weak deflection limit, and the full gravitational deflection angle (including all order perturbation contributions applicable to both weak and strong deflection limits) is also calculated numerically as comparisons. It enables us to analyze the perfect fluid dark matter influences on gravitational deflection angle and gravitational lensing beyond the leading order, which were not sufficiently studied in previous works. Assuming $M \sim \lambda_{\text{DM}} \sim Q$, our results show that dark matter can greatly influence the gravitational lensing of central black holes.
	
\ \ 


\ \ \ \ \ \ \ \ \ \ \ \ \ \ \   Dark Matter Halo}

\begin{document}
\maketitle
\flushbottom

\section{Introduction \label{sec:1}}

The study of black hole spacetimes is an important topic in general relativity and other gravity theories. Significantly important information on gravitation, thermodynamics and quantum effects in curved spacetime can be revealed through black holes \cite{Bekenstein1973,Hawking1975,Unruh1981,Ryu2006}. In the past several years, a number of progresses on theoretical and observational explorations of black holes have been witnessed. The gravitational wave signals were detected by LIGO and Virgo from the merging of binary black holes \cite{LIGO2016}. The high resolution black hole images at the center of M87 and our galaxy were successfully observed by Event Horizon Telescope \cite{EHT2019,EHT2022}.

Dark matter is one of the most attractive topics among elementary particle physics, astrophysics, astronomy and cosmology. The accumulating evidences for the existence of dark matter have been found from the rotation curves of spiral galaxies \cite{Rubin1970,Corbelli2000}, gravitational lensing \cite{Clowe2006}, large scale structure formation of the universe \cite{Davis1985}, cosmic microwave background and baryon acoustic oscillations \cite{WMAP2011,Planck2014}. The combined observations from cosmic microwave background indicate that $26.8\%$ of our universe is made up of dark patter, and $68.3\%$ of our universe is made up of dark energy \cite{Planck2014}. The dark matter could affect the evolution of our universe, and it determines the fate of our universe. The most favorable candidates for dark matter are some unknown particles predicted by theories beyond the Standard Model, such as weakly interacting massive particles (WIMPs), axions, sterile neutrinos, dark photons, millicharged particles \cite{Bertone2005,Boehm2004,Feng2009,Graham2015,Schumann2019,Boyarsky2019,Qiao2021a}. This dark matter usually forms halo structures in large number of galaxies \cite{Navarro1996,Navarro1997,Cooray2002}. In this way, the supermassive black holes in the center of galaxies are surrounded by these dark matter halos. Therefore, studying and analyzing the effects of dark matter halo on supermassive black holes in the galaxy center are of great significance. Recently, a number of works have been emerged to study the dark matter effects on black hole behaviors, such as the circular geodesics \cite{Das2021,Rayimbaev2021}, black hole shadow \cite{HouX2018a,HouX2018b,Konoplya2019,Jusufi2020,Saurabh2021,Das2022,MaSJ2022,Pantig2022b,Ghosh2023,Xavier2023}, gravitational lensing \cite{Haroon2019,Islam2020,Pantig2020,Pantig2021,Pantig2022,Pantig2022c,Atamurotov2021,Atamurotov2022}, accretion disk \cite{Fard2022}, thermodynamics \cite{XuZ2019,TaoJ2021,TaoJ2022}, quasi-normal mode \cite{Jusufi2020b,LiuD2022}, black hole echoes \cite{LiuD2021} and black hole merging process \cite{Bamber2022}.

The properties of dark matter (as well as its halo structures in galaxies) could be investigated in a number of observational methods. The gravitational lensing is one of the most important methods. In the past few decades, it has become a valuable tool in astronomy and cosmology to test gravity theories and astrophysical models \cite{Wambsganss1998,Bartelmann2001,Mollerach2002,Uitert2012}. Particularly, recent studies suggest that the dark matter distribution in galaxies can be revealed from gravitational lensing observations \cite{Clowe2006,Uitert2012,Brimioulle2013}. The amounts of dark matter and the properties of dark matter halos in galaxies and galaxies clusters are strongly constrained by gravitational lensing observations.

In this work, we aim to get a comprehensive study of the dark matter effects on the gravitational lensing for supermassive black holes in the galaxy center. We choose two classical renowned black holes --- the Schwarzschild black hole and the charged Reissner-Nordstr\"om (RN) black hole --- to carry out the calculations. As the most popular and representative examples in general relativity and gravity theories, the Schwarzschild and charged RN black holes could reveal most of universal properties of the more complex spherically symmetric black holes. Further, we assume that the dark matter, which surrounds the central black hole, can be effectively modeled by a perfect fluid with energy-momentum tensor given by $T_{\mu\nu} = (\rho+p)u_{\mu}u_{\nu} + p g_{\mu\nu}$. This kind of dark matter is usually called the perfect fluid dark matter (PFDM) in literature \cite{Rahaman2010}. Since 2019, several studies on gravitational deflection angle for black holes in the presence of PFDM have been emerged \cite{Haroon2019,Atamurotov2021,Atamurotov2022,YPHu2023}. Haroon \emph{et al.} and Atamurotov \emph{et al.} derived the leading order gravitational deflection angle for Schwarzschild and charged black holes in PFDM \cite{Haroon2019,Atamurotov2022}. The second order contributions from dark matter are not sufficiently studied in these works. Very recently, Gao \emph{et al.} gave a second order deflection angle for charged black hole immersed in PFDM by applying the ``thin dark matter'' assumption (the mass of black hole is much larger than the dark matter) in the weak deflection limit \cite{YPHu2023}. Exploring the higher order contributions (at least the second order) to gravitational deflection angle for these black holes in PFDM and studying their influences on gravitational lensing observations in the general case beyond the ``thin dark matter'' is very important, and this is one of the purposes of our work. In the present work, we assume that the black hole mass, black hole charge, and dark matter parameter are in the same order of magnitude ($M \sim Q \sim \lambda_{\text{DM}}$). The gravitational deflection angle of light, photon sphere, black hole shadow radius, gravitational lens equation and Einstein ring for Schwarzschild and charge black holes surrounded by PFDM are calculated analytically and numerically. We mainly focus on the dark matter effects on central black holes in gravitational deflection and gravitational lensing. Specifically, a second order analytical expansion of gravitational deflection angle is obtained in the weak deflection limit, and the full gravitational deflection angle (including all order perturbation contributions applicable to both weak and strong deflection limits) is also calculated numerically for comparison. Hopefully, the conclusions generated from these classical black hole could give us hints on the gravitational lensng of more complex black holes surrounded by dark matter halos.

In the present work, we use two approaches to calculate the gravitational deflection angle of light. One is a geometric approach developed by G. W. Gibbons and M. C. Werner \cite{Gibbons2008}, in which the gravitational deflection angle is calculated by applying Gauss-Bonnet theorem in differential geometry and topology. The other one is the conventional geodesic approach in gravity theories, in which the gravitational deflection angle is calculated by solving the trajectories of particle orbits (which are null geodesics for photon orbits). In our work, the second-order expansion of gravitational deflection angle is obtained analytically using the Gibbons and Werner approach, and the full gravitational deflection angle with all order perturbation contributions included is calculated numerically using the geodesic approach. In the calculation of photon spheres and black hole shadow radius, we resort to the effective potential of test particle moving in the gravitational field. Besides, we can also use a geometric approach developed by Qiao \emph{et al.} \cite{Qiao2022a,Qiao2022b,Cunha2022}, in which the stable and unstable photon spheres are determined using the Gaussian curvature and geodesic curvature. This approach could get completely equivalent results to the conventional approach (based on the effective potential of test particles). Furthermore, in the calculation of Einstein ring angles, we numerically solve the gravitational lens equation in weak deflection limit.

This paper is organized in the following way. Section \ref{sec:1} gives the motivations and background introductions of this work. In section \ref{sec:2}, the Schwarzschild black hole and the charged RN black hole immersed in perfect fluid dark matter (PFDM) halo are briefly reviewed. Section \ref{sec:3} gives descriptions of the theoretical framework in this work. The approaches to calculate the gravitational deflection angle, photon sphere, black hole shadow radius, lens equation and Einstein ring angle are presented systematically in this section. Section \ref{sec:4} presents the analytical and numerical results on gravitational deflection angle of light. Section \ref{sec:5} gives the numerical results on photon spheres and black hole shadows. Section \ref{sec:6} briefly discusses the observable in gravitational lensing observations, including the lens equation and Einstein ring angle. The summary and conclusions are given in section \ref{sec:7}. Furthermore, in this work, the natural unit $G=c=1$ is adopted.

\section{Schwarzschild and Charged Reissner-Nordstr\"om (RN) Black Holes Immersed in Perfect Fluid Dark Matter (PFDM) Halo \label{sec:2}} 

In this section, we give an introduction of the spacetime metric for black holes immersed in dark matter halos. To simply model the supermassive black hole in the galaxy center surrounded by dark matter, in the present work, we consider two classical black holes --- the Schwarzschild black hole and charged Reissner-Nordstr\"om (RN) black hole --- immersed in dark matter halos. As typical examples in general relativity and gravity theories, the Schwarzschild and charged RN black holes could reflect most of universal properties for the more complex spherically symmetric black holes. 

The spacetimes generated by Schwarzschild and charged RN black holes are spherically symmetric. Furthermore, modern astrophysical observations suggest that dark matter halos in many galaxies are nearly spherically symmetric. Particularly, a number of spherically
symmetric dark matter mass density profile (such as the Einasto, NFW, Burkert, Brownstein, Moore models) were proposed from the observations and simulations \cite{Navarro1996,Navarro1997,Burkert1995,Moore1998,Brownstein2006,Dutton2014}. In recent years, these spherically symmetric dark matter halo models successfully gave consistent results with the observed flat rotational curves as well as other large scale observations, and constrained the amount of dark matter in our galaxy and other galaxies \cite{Dutton2014,Pato2015,Salucci2019,Sofue2020}. Therefore, for simplicity, we can use the static and spherically symmetric spacetime to give a description of the supermassive black hole immersed in a dark matter halo.
\begin{eqnarray}
	d\tau^{2} & = & g_{\mu\nu}dx^{\mu}dx^{\nu} 
	 =  f(r)dt^{2}
	-\frac{1}{f(r)}dr^{2}
	-r^{2}
	(d\theta^{2}+\sin^{2}\theta d\phi^{2}) \label{spacetime metric}
\end{eqnarray}
the function $f(r)$ is defined through the black hole mass $M$, black hole charge $Q$ and contributions from dark matter halo:
\begin{equation}
	f(r) = 1 - \frac{2M}{r} + \frac{Q^{2}}{r^{2}} 
	+ \text{contributions from dark matter}
\end{equation}
For such spherically symmetric spacetime, the modulus of tangent vector gives the rotational velocity for test particles moving along circular orbits in the equatorial plane \cite{Matos2000,Xu2018,Jusufi2020}
\begin{equation}
	v_{\text{tangent}}^{2} = \frac{r}{\sqrt{f(r)}}\cdot\frac{d \sqrt{f(r)}}{dr}
	= r\cdot\frac{d \ln\sqrt{f(r)}}{dr} \label{tangent volocity}
\end{equation}
The detailed analytical expression of function $f(r)$ depends on the dark matter models. Different dark matter models may give different  functions $f(r)$, which must be able to give the flat rotational curves observed in many galaxies \cite{Rubin1970,Corbelli2000}. In the present work, to give a concrete study of the dark matter effects on gravitational deflection and gravitational lensing, we consider the dark matter described by a perfect fluid with energy-momentum tensor
\begin{equation}
	T_{\mu\nu}=(\rho+p)u_{\mu}u_{\nu}+p g_{\mu\nu}
\end{equation}
It is called the perfect fluid dark matter (PFDM) in literature \cite{Rahaman2010}.

In this work, in order to effectively study the behavior of central supermassive black holes in dark matter halos in a simpler way, we consider the Schwarzschild and charged RN black holes surrounded by a PFDM halo. These black hole solutions (in the presence of dark matter) have been studied extensively in the literature, and the effective spacetime metrics have been successfully derived \cite{LiMH2012,Heydarzade2017,XuZ2019,Jusufi2020b,Haroon2019,Das2021}. Particularly, for the Schwarzschild black hole immersed in PFDM, the function $f(r)$ can be expressed as:
\begin{equation}
	f(r)=1-\frac{2M}{r} 
	+ \frac{\lambda_{\text{DM}}}{r} \cdot  \ln\bigg(\frac{r}{|\lambda_{\text{DM}}|}\bigg)
	\label{spacetime metric - perfect fluid dark matter}
\end{equation}
For the charged RN black hole immersed in PFDM, the function $f(r)$ can be expressed as:
\begin{equation}
	f(r) = 1 - \frac{2M}{r} + \frac{Q^{2}}{r^{2}}
	+ \frac{\lambda_{\text{DM}}}{r} \cdot \ln\bigg(\frac{r}{|\lambda_{\text{DM}}|}\bigg)
	\label{RN spacetime metric - perfect fluid dark matter}
\end{equation}
Here, $M$ and $Q$ are the mass and charge of black hole. The $\lambda_{\text{DM}}$ is the dark matter parameter, which is proportional to the mass density of PFDM.

The above spacetime metric (\ref{spacetime metric - perfect fluid dark matter}) and (\ref{RN spacetime metric - perfect fluid dark matter}) are derived from the Lagrangian 
\begin{eqnarray}
	\mathcal{S} & = & \int d^{4}x\sqrt{-g} \cdot \mathcal{L} 
	 =  \int d^{4}x\sqrt{-g} \cdot 
	\bigg[
	\frac{1}{16\pi G} R 
	+ \frac{1}{4} F^{\mu\nu}F_{\mu\nu}
	+ \mathcal{L}_{\text{DM}}
	\bigg]
\end{eqnarray}
with the energy-momentum tensor of dark matter to be $T_{\mu\nu}=(\rho+p)u_{\mu}u_{\nu}+p g_{\mu\nu}$. Furthermore, the mass density of PFDM can be determined by the dark matter parameter $\lambda_{\text{DM}}$ via \cite{Das2021}
\begin{equation}
	\rho = \frac{\lambda_{\text{DM}}}{8\pi r^{3}}
\end{equation}
In the PFDM model, to be consistent with the observed flat rotational curves in galaxies, the rotational velocity and dark matter parameter should satisfy  \cite{LiMH2012,Lobo2008,Saurabh2021}
\begin{equation}
	v^{2} = v_{\text{tangent}}^{2} 
	\to \frac{\lambda_{\text{DM}}}{2r_{0}}
\end{equation}
where $v_{\text{tangent}}$ is the modulus of tangent vector for stars traveling along circular orbits, and $r_{0}$ is a characteristic scale in the galaxy. 
In a typical galaxy, when we discussing the gravitational deflection angle and gravitational lensing, it is reasonable to have the following relation \cite{LiMH2012}
\begin{equation}
	r_{H} \ll b \ll r_{0} < r_{\text{halo}}
\end{equation}
where $r_{H}$ is the horizon radius of central supermassive black hole, $r_{\text{halo}}$ denotes the scale of dark matter halo, and $b$ is the impact parameter for light rays in gravitational lensing observations. Furthermore, the 
dark matter can contribute to a large amount of mass in a galaxy. In this work,  we assume that the black hole mass, black hole charge and dark matter parameter are in the same order of magnitude, namely $M \sim Q \sim \lambda_{\text{DM}}$ 
\footnote{In most of the galaxies, the electric charge of central supermassive black hole is much smaller than the black hole mass ($Q \ll M$). However, to emphasize the effects of charged black holes in the presence of PFDM, we assume the black hole charge and black hole mass are in the same order of magnitude in the analytical calculation of gravitational deflection angle.}. 
Under these assumptions, it is convenient to introduce small parameters in the analytical expansion of gravitational deflection angle
\begin{equation}
	\epsilon \sim \frac{M}{b} \sim \frac{\lambda_{\text{DM}}}{b} \sim \frac{Q}{b} \sim \frac{r_{H}}{b} \ll 1
	\ \ \ \ \ 
	\eta \sim \frac{\ln\epsilon^{-1}}{\epsilon^{-1}}=\epsilon\cdot\ln\frac{1}{\epsilon} \ll 1
\end{equation} 

\section{Theoretical Framework \label{sec:3}}

This section gives an introduction on the theoretical framework of our work. The subsection \ref{sec:3a} and subsection \ref{sec:3b} describe two approaches in calculating the gravitational defection angle. Subsection \ref{sec:3a} introduces the Gibbons and Werner approach, in which the gravitational deflection angle is calculated using the Gauss-Bonnet theorem. Subsection \ref{sec:3b} introduces the conventional geodesic approach, in which the gravitational deflection angle is obtained by solving the trajectories of null geodesics. Subsection \ref{sec:3c} discusses the photon spheres near black holes and the shadow radius detected by an observer. Subsection \ref{sec:3d} is devoted to the lens equation and Einstein rings in gravitational lensing observations.

\subsection{Gibbons and Werner Approach (Gauss-Bonnet Theorem) to Gravitational Deflection Angle \label{sec:3a}}

According to Einstein's general theory of relativity, the gravitational interaction can be understood as a geometric effect. It is possible to analyze the gravitational deflection and gravitational lensing using new techniques from geometry and topology. Recently, an approach calculating the gravitational deflection angle of particles utilizing the Gauss-Bonnet theorem emerged. This approach was first introduced in a pioneering work given by G. W. Gibbons and M. C. Werner \cite{Gibbons2008}, in which a topological interpretation on the gravitational deflection was given. In later years, this approach has been successfully applied to large number of gravitational systems, and consistent results with conventional geodesic approach have been obtained in the weak deflection limit \cite{Werner2012,Ishihara2016a,Ishihara2016b,Ovgun2018,Jusufi2018,Crisnejo2018a,Crisnejo2018b,Crisnejo2019,Javed2019,Li2020a,Li2020b,Takizawa2020,Qiao2021b,Liu2021,Zhang2021,Kumaran2021,Atamurotov2021c,Jha2022,Parbin2022,Huang2022a,Huang2022b,LiuLH2022}. 

The Gauss-Bonnet theorem is one of the most profound theorem in differential geometry and topology. It provide nontrivial connections between curvatures and topological invariant in a curved manifold. In a two dimensional curved manifold, the mathematical description of Gauss-Bonnet theorem is \cite{Chern}
\begin{equation}
	\int_{D}\mathcal{K} dS + \int_{\partial D}\kappa_{g} dl + \sum_{i=1}^{N}\theta_{i} = 2 \pi \chi(D) 
	\label{Gauss-Bonnet theorem}
\end{equation}
Here, $D$ is a region in curved manifold, $\mathcal{K}$ is the Gaussian curvature, $\kappa_{g}$ is the geodesic curvature of boundary $\partial D$, $\chi(D)$ is the Euler characteristic number for region $D$, and $\theta_{i}$ is the exterior angle for each discontinuous point of boundary $\partial D$. 

In Gibbons and Werner approach, the gravitational deflection angle of light is calculated by applying the Gauss-Bonnet theorem in the optical geometry of black hole spacetime. For static and stationary black holes, the optical geometry can be obtained from the four dimensional black hole spacetime by imposing the null constraint $d\tau^{2}=0$ \cite{Gibbons2008,Gibbons2009,Werner2012}
\begin{equation}
	\underbrace{d\tau^{2} = g_{\mu\nu}dx^{\mu}dx^{\nu}}_{\text{Spacetime Geometry}}
	\ \ \overset{d\tau^{2}=0}{\Longrightarrow} \ \ 
	\underbrace{dt^{2} = g^{\text{OP}}_{ij}dx^{i}dx^{j}}_{\text{Optical Geometry}}
	\label{optical geometry0}
\end{equation} 
This optical geometry is motivated by the generalization of the famous Fermat’s principle to a curved spacetime. The null
geodesic curve $\gamma=\gamma(\lambda)$ in the spacetime geometry $d\tau^{2}=g_{\mu\nu}dx^{\mu}dx^{\nu}$ becomes a spatial geodesic in the optical geometry $dt^{2}=g^{\text{OP}}_{ij}dx^{i}dx^{j}$ \footnote{The classical Fermat’s principle states that: light rays always travel along particular spatial curves such that the optical distance $s_{ab}^{\text{OP}}=\int_{a}^{b} n(x)dx$ is minimal (where $n(x)$ is the reflective index). The classical Fermat’s principle is restricted in a flat space. Fortunately, this principle can be generalized to a static or stationary curved spacetime. In the static or stationary curved spacetime, one can always have a global choice of the time coordinate $t$, massless photons (which travel along null geodesics in spacetime geometry) starting from a fixed emission point at a given time $t_{a}$ would eventually make the arrival time $t_{b}$ minimal, which means the variational condition $\delta\big[\int_{a}^{b}dt\big]=0$ is satisfied. In this way, the photon orbits must be spatial geodesics in the optical geometry. In the optical geometry, the optical metric $dt^{2}=g_{ij}^{\text{OP}}dx^{i}dx^{j}$ gives the infinitesimal change of time square, and the spatial geodesics in optical geometry always make the variation $\delta\big[\int_{a}^{b}dt\big]=0$. In the generalization of Fermat's principle to static or stationary curved spacetime, the spatial length in optical geometry $l_{ab}^{OP}=\int_{a}^{b}dt=\int_{a}^{b}\sqrt{g_{ij}^{OP}dx^{i}dx^{j}}$ effectively plays the role of ``optical distance''. The discussion of the generalized Fermat's principle in a static or stationary curved spacetime can be found in references \cite{Werner2012,Jusufi2018,Gibbons2009,Landau1975,Perlick2000,LinQ2008}.}. 
Particularly, if we consider a four dimensional spherically symmetric black hole spacetime 
\begin{eqnarray}
	d\tau^{2} & = & g_{\mu\nu}dx^{\mu}dx^{\nu} 
	 =  f(r)dt^{2}
	-\frac{1}{f(r)}dr^{2}
	-r^{2}
	(d\theta^{2}+\sin^{2}\theta d\phi^{2}) \label{spacetime}
\end{eqnarray}
its optical geometry gives a three dimensional Riemannian manifold \cite{Gibbons2008,Gibbons2009}
\begin{eqnarray}
	dt^{2} & = & g^{\text{OP}}_{ij}dx^{i}dx^{j} 
	 = \frac{1}{f(r)}
	\bigg[
	\frac{1}{f(r)}dr^{2} +r^{2}d\theta^{2}
	+r^{2} \sin^{2}\theta d\phi^2 
	\bigg]
\end{eqnarray}
Furthermore, for spherically symmetric black holes, without loss of generality, we can restrict the above optical geometry (as well as the optical geometry metric $g^{\text{OP}}_{ij}$) in the equatorial plane. In this way, a two dimensional Riemannian manifold is eventually obtained, and the application of Gauss-Bonnet theorem in equation (\ref{Gauss-Bonnet theorem}) becomes available.

\begin{figure}
	\includegraphics[width=0.725\textwidth]{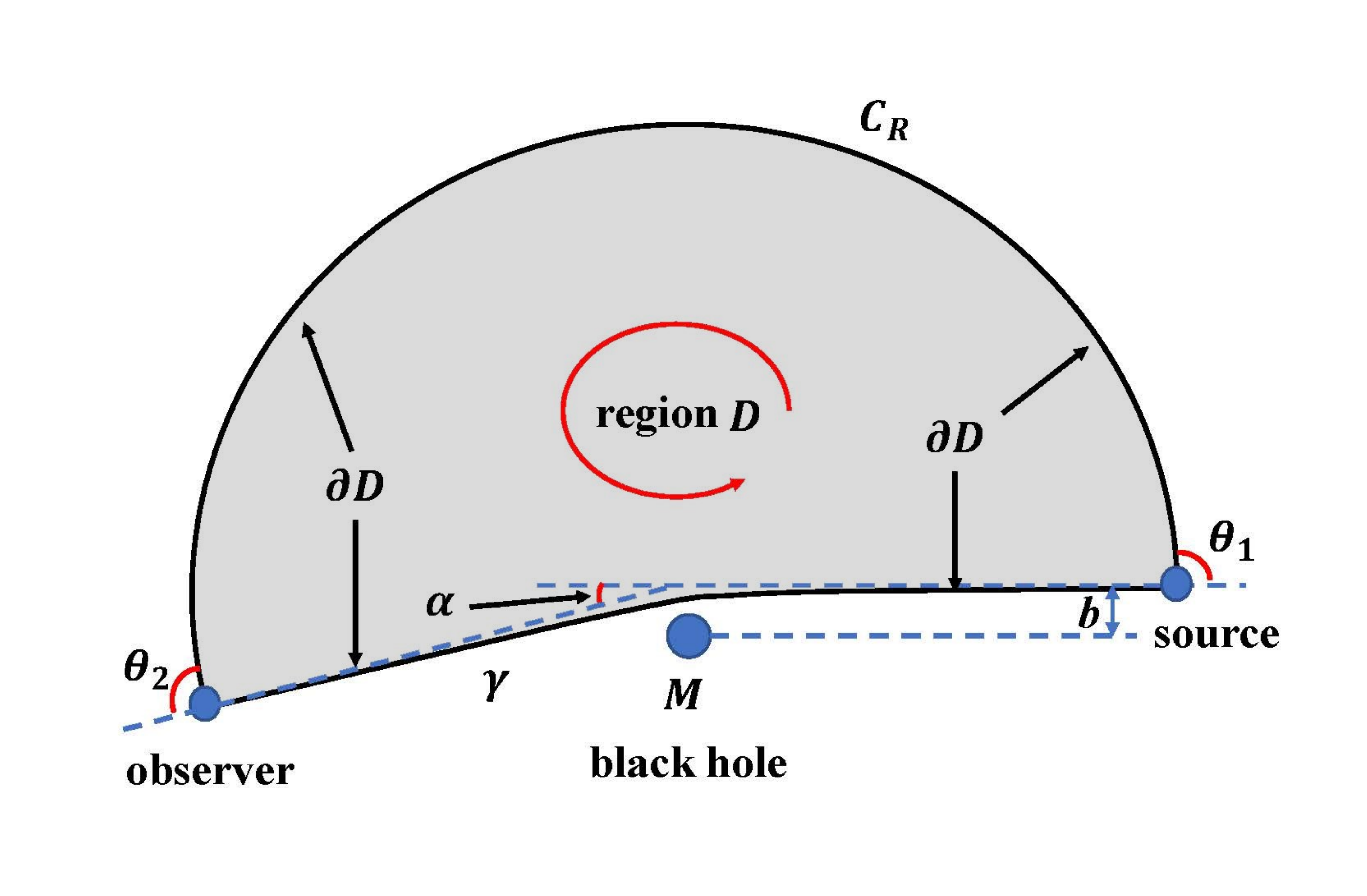}
	\caption{This figure shows the choice of region $D$ in the equatorial plane of optical geometry for asymptotically flat black holes. The boundary $\partial D$ consists of two parts: a photon orbit $\gamma$ from light source to observer, and a circular arc connect with the source and observer. Furthermore, the direction of boundary $\partial D$ in the contour integral $\int_{\partial D} \kappa_{g} dl$ is chosen to be counterclockwise. Note that for the photon orbit $\gamma$, this choice of direction (which is from observer to light source) is opposite to the propagation of photon beams.}
	\label{figure1}
\end{figure}

In the calculations of gravitational deflection angles, the region $D$ in Gauss-Bonnet theorem must be picked in the equatorial plane of optical geometry. For an asymptotically flat black hole, when light source and observer are both very far from the central supermassive black hole, $D$ is usually chosen to be simply connected such that the light source and observer are connected by its boundary $\partial D$, as shown in figure \ref{figure1}. The central black hole is located outside the region $D$, so that spacetime singularities are excluded from this region. For this picked region $D$, the exterior angles for discontinuous points of boundary $\partial D$ are approximated as
\begin{equation}
	\theta_{1} \approx \theta_{2} \approx \frac{\pi}{2}
	\ \ \Rightarrow \ \ 
	\sum_{i=1}^{N}\theta_{i} = \theta_{1} + \theta_{2} \approx \pi \label{exterior angle}
\end{equation}
The Euler characteristic number for the simply connected region $D$ is
\begin{equation}
	\chi(D)=1 \label{Eular characteristic number}
\end{equation}
Furthermore, for asymptotically flat black hole spacetimes, many studies have shown that the integration of geodesic curvature $\kappa_{g}$ along the circular arc $C_{R}$ in the $R \to \infty$ limit reduces to \cite{Javed2019,Kumaran2021}
\begin{equation}
	\int_{\partial D}\kappa_{g} dl = \lim_{R \to \infty} \int_{C_{R}}\kappa_{g}(C_{R}) dl
	\approx \pi+\alpha
	\label{geodesic curvature contour integral}
\end{equation}
where $\alpha$ is the gravitational deflection angle of light.

Combining the exterior angles in equation (\ref{exterior angle}), Euler characteristic number in equation (\ref{Eular characteristic number}) and contour integral of geodesic curvature in equation (\ref{geodesic curvature contour integral}), the Gauss-Bonnet theorem in equation (\ref{Gauss-Bonnet theorem}) leads to
\begin{equation}
	\int_{D}\mathcal{K} dS + \int_{\partial D}\kappa_{g} dl + \sum_{i=1}^{N}\theta_{i} 
	= \int_{D}\mathcal{K} dS+ (\pi+\alpha) + \pi 
	= 2 \pi \chi(D) = 2 \pi
\end{equation}
Eventually, the gravitational deflection angle of light can be calculated through the integration of Gaussian curvature over the picked region $D$
\begin{equation}
	\alpha = -\int_{D}\mathcal{K} dS
	\label{Gauss-Bonnet gravitational deflection}
\end{equation}
From this approach, the gravitational deflection angle is directly linked with geometric and topological properties of optical geometry of black holes. It can give us new insights on gravitational deflection and gravitational lensing.

\subsection{Conventional Geodesic Approach to Gravitational Deflection Angle \label{sec:3b}}

In the conventional geodesic approach, the gravitational deflection angle is calculated by solving the differential equations of null geodesics. In the past few decades, this approach has been tested by large numbers of observations, and it has become a widely adopted approach in physics and astronomy. This approach can effectively duel with the gravitational lensing in both strong and weak deflection limits \cite{Carroll2019,Weinberg1972,Virbhadra1998,Virbhadra2000,Virbhadra2002,Eiroa2002,Keeton2005,Iyer2007,Bozza2008,Kogan2008,Virbhadra2009,Kim2021,Tsukamoto2021,Perlick2022b,Tsukamoto2022,Atamurotov2021b,Atamurotov2021d,Atamurotov2022a,LiuLH2022b}. 

The basic idea of the conventional geodesic approach is solving the differential equations of null geodesics and computing the variation of azimuthal angle $\phi$ in the photon trajectory. For a spherically symmetric spacetime
\begin{equation}
	d\tau^{2} = f(r)dt^{2} -\frac{1}{f(r)}dr^{2} 
	-r^{2}(d\theta^{2}+\sin^{2}\theta d\phi^{2})
	\label{spacetime metric 3}
\end{equation}
the following conserved quantities can be introduced \cite{Carroll2019,Weinberg1972}
\begin{subequations}
	\begin{eqnarray}
		J & \equiv & r^{2}\sin^{2}\theta \frac{d\phi}{d\lambda}
		\\
		E & \equiv &  f(r)\frac{dt}{d\lambda}
		\\
		\epsilon & \equiv & g_{\mu\nu}dx^{\mu}dx^{\nu} 
		= f(r) \bigg( \frac{dt}{d\lambda} \bigg)^2
		- \frac{1}{f(r)} \bigg( \frac{dr}{d\lambda} \bigg)^{2}
		- r^{2} \bigg( \frac{d\theta}{d\lambda} \bigg)^{2} 
		- r^{2}\sin^{2}\theta \bigg( \frac{d\phi}{d\lambda} \bigg)^{2}
	\end{eqnarray}
\end{subequations}
Here, $\lambda$ is an affine parameter, $J$ is the conserved angular momentum along a particle orbit, and $E^{2}/2$ can be viewed as the conserved energy along a particle orbit. For text particles moving in the equatorial plane $\theta=\pi/2$, the following reduced differential equations can be obtained using these conserved quantities
\begin{subequations}
	\begin{eqnarray}
		&& \frac{1}{2} \bigg( \frac{dr}{d\lambda} \bigg)^{2} + \frac{1}{2} f(r) \bigg[ \frac{J^{2}}{r^{2}} + \epsilon \bigg]
		= \frac{1}{2} \bigg( \frac{dr}{d\lambda} \bigg)^{2} + V_{\text{eff}}(r)
		= \frac{1}{2}E^{2} \label{reduced differential equation1} 
		\\
		&& \bigg( \frac{dr}{d\phi} \bigg)^{2}
		= \frac{r^{4}}{b^{2}} - r^{4} \bigg[ \frac{f(r)}{r^{2}} + \frac{\epsilon \cdot f(r)}{J^{2}} \bigg] 
		\label{reduced differential equation2}
	\end{eqnarray}
\end{subequations}
Here, $V_{\text{eff}}(r) = \frac{f(r)}{2} \big[ \frac{J^{2}}{r^{2}} + \epsilon \big]$ is the effective potential of test particles moving in the spherically symmetric gravitational field, and the impact parameter is defined as $b\equiv |J/E|$. For massless (or massive) particles traveling along null (or timelike) geodesics, the quantity $\epsilon$ takes the value $\epsilon=0$ (or $\epsilon=1$).

From the reduced differential equation (\ref{reduced differential equation2}), one can get
\begin{equation}
	\frac{dr}{d\phi} = \pm r^{2} \sqrt{\frac{1}{b^{2}}-\bigg[\frac{f(r)}{r^{2}}+\frac{\epsilon \cdot f(r)}{J^{2}}\bigg]}
\end{equation}
The plus and minus sign $\pm$ can be determined in the following way. When a test particle moves along the scattering orbit from the infinity, the radial coordinate $r$ decreases and the azimuthal angle $\phi$ increases, until this particle reaches the closest distance $r_0$ to central supermassive black hole. After the test particle passing the turning point $r=r_{0}$, its radial coordinate $r$ starts to increase as the azimuthal angle $\phi$ increases. In this way, we have the relations
\begin{subequations}
	\begin{eqnarray}
		\frac{dr}{d\phi} & = & - r^{2} \sqrt{\frac{1}{b^{2}}-\bigg[\frac{f(r)}{r^{2}}+\frac{\epsilon \cdot f(r)}{J^{2}}\bigg]} < 0 \ \ \ \text{particle moving from $r=\infty$ to the tuning point $r=r_{0}$} 
		\ \ \ \ \ \ \ \ \ \ 
		\\
		\frac{dr}{d\phi} & = & r^{2} \sqrt{\frac{1}{b^{2}}-\bigg[\frac{f(r)}{r^{2}}+\frac{\epsilon \cdot f(r)}{J^{2}}\bigg]} > 0 \ \ \ \ \text{particle moving from the tuning point $r=r_{0}$ to $r=\infty$}  
		\ \ \ \ \ \ \ \ \ 
	\end{eqnarray}
\end{subequations} 

In the gravitational lensing, when the light source and observer are both located at infinity, the gravitational deflection angle of light is just twice of the variation $\Delta\phi$ as radial coordinate $r$ changes from $r=r_{0}$ to $r=\infty$ \cite{Weinberg1972,Virbhadra1998}.
\begin{eqnarray}
	\alpha & = & 2|\phi(\infty)-\phi(r_{0})|-\pi \nonumber
	\\
	& = & 2\bigg| 
	\int_{r_{0}}^{\infty} \frac{d\phi}{dr}dr 
	\bigg|
	- \pi 
	 =  2\int_{r_{0}}^{\infty}
	\frac{dr}{r^{2}\sqrt{\frac{1}{b^{2}}-\frac{f(r)}{r^{2}}}}
	-\pi 
	\label{gravitational deflection angle from null geodesic}
\end{eqnarray}
Here, $\epsilon=0$ has been used for massless photons. From this expression, it is obvious that the gravitational deflection angle $\alpha$ increases monotonically as the minimal distance $r_{0}$ decreases. In the integration process, the turning point $r=r_{0}$ must be solved. For the closet distance to central black hole, the derivative $dr/d\lambda$ in equation (\ref{reduced differential equation1}) vanishes automatically
\begin{eqnarray}
	\frac{dr}{d\lambda}\bigg|_{r=r_{0}} = 0 
	\ & \Rightarrow & \
	\frac{1}{2}f(r_{0}) \frac{J^{2}}{r_{0}^{2}} = \frac{1}{2}E^{2} \nonumber
	\\
	\ & \Rightarrow & \ b^{2} = \frac{J^{2}}{E^{2}} = \frac{r_{0}^{2}}{f(r_{0})}
\end{eqnarray}
Given the impact parameter $b$ for a photon orbit, the closet distance $r_{0}$ can be solved from this equation.

In the numerical calculation, it is useful to find other expressions equivalent to equation (\ref{gravitational deflection angle from null geodesic}) to prevent the integration variable reaching $r=\infty$. One simple way is adopting $u=\frac{1}{r}$ to be the integration variable
\begin{equation}
	\alpha = 2\int_{0}^{u_{0}} \frac{du}{\sqrt{\frac{1}{b^{2}}-f(u) \cdot u^{2}}}
	-\pi 
\end{equation}
with upper limit to be $u_{0}=\frac{1}{r_{0}}$. Another way is assigning $x=\frac{r_{0}}{r}$ to be the integration variable
\begin{eqnarray}
	\alpha & = & 2\int_{r_{0}}^{\infty}
	\frac{dr}{r^{2}\sqrt{\frac{1}{b^{2}}-\frac{f(r)}{r^{2}}}}
	-\pi 
	 =  2\int_{r_{0}}^{\infty} \frac{r_{0} dr}{r^{2}\sqrt{f(r_{0})-\big(\frac{r_{0}}{r}\big)^{2}\cdot f(r)}} -\pi \nonumber
	\\
	& = & 2\int_{0}^{1}\frac{dx}{\sqrt{F_{0}-F(x)}} -\pi
	\label{gravitational deflection angle from null geodesic --- equivalent expression}
\end{eqnarray}
where $F_{0}$ and $F(x)$ are defined through
\begin{subequations}
	\begin{eqnarray}
		F(x) & \equiv & x^{2}\cdot f(r) 
		= x^{2}\cdot f\bigg(\frac{r_{0}}{x}\bigg)
		\\
		F_{0} & \equiv & f(r_{0}) = F(x=1)
	\end{eqnarray}
\end{subequations}
In the presented work, we choose the equation (\ref{gravitational deflection angle from null geodesic --- equivalent expression}) to carry out the numerical calculations.

\subsection{Photon Sphere and Black Hole Shadow \label{sec:3c}}

The photon spheres and black hole shadows are crucial quantities in the gravitational lensing observations. These quantities could reflect many intrinsic properties of black holes. Recently, the black hole shadow images in the center M87 galaxy and Sgr A* caught by Event Horizon Telescope (EHT) have stimulated large numbers of investigations on black holes.

Conventionally, the exact positions of photon spheres near black hole are achieved using the effective potentials of photons \cite{Carroll2019,Hartle2021,Pugliese2011,Johannsen2013,Gan2021,Gan2021b,Atamurotov2020a,Atamurotov2021a,Perlick2022,WangMZ2022,Mustafa2022,GuoGZ2022a}. For static and spherically symmetric black holes described in equation (\ref{spacetime metric 3}), the photon equation of motion has a simple form 
\begin{equation}
	\frac{1}{2}\cdot\frac{d^{2}r}{d\lambda^{2}}+V_{\text{eff}}(r)=\frac{1}{2}\cdot E^{2}
\end{equation}
with the photon effective potential to be $ V_{\text{eff}}(r)=\frac{f(r)}{2}\cdot\frac{J^{2}}{r^{2}} $.
The local maximal points of effective potential give the unstable photon spheres, while the local minimal points of effective potential give the stable photon spheres near black holes. 
\begin{subequations}
\begin{eqnarray}
	\frac{dV_{\text{eff}}(r)}{dr}=0 & \Leftrightarrow & \text{photon sphere} 
	\\
	\frac{dV_{\text{eff}}(r)}{dr}=0 \ \ \text{and} \ \  \frac{d^{2}V_{\text{eff}}(r)}{dr^{2}}<0 & \Leftrightarrow & \text{unstable photon sphere} 
	\\
	\frac{dV_{\text{eff}}(r)}{dr}=0 \ \ \text{and} \ \  \frac{d^{2}V_{\text{eff}}(r)}{dr^{2}}>0 & \Leftrightarrow & \text{stable photon sphere} 
\end{eqnarray}
\end{subequations}

Recently, a new geometric approach is proposed such that the stable and unstable photon spheres are determined by the Gaussian curvature and geodesic curvature introduced in subsection \ref{sec:3a} \cite{Qiao2022a,Qiao2022b}. In this geometric approach, the photon spheres are determined by circles in the equatorial plane of optical geometry with zero geodesic curvature $\kappa_{g}(r)=0$. In addition, the negative Gaussian curvature indicates that the corresponding photon spheres are unstable (with $\kappa_{g}(r)=0$ and $\mathcal{K}(r)<0$), while the positive Gaussian curvature indicates that the photon spheres are stable (with $\kappa_{g}(r)=0$ and $\mathcal{K}(r)>0$). Furthermore, it is also proved that this geometric approach is completely equivalent to the conventional approach using the effective potential of photons.
\begin{subequations}
	\begin{eqnarray}
		\frac{dV_{\text{eff}}(r)}{dr} \bigg|_{r=r_{ph}} = 0 
		\ & \Leftrightarrow & \  
		\kappa_{g}(r=r_{ph})=0 
		\\
		\frac{d^{2}V_{\text{eff}}(r)}{dr^{2}} \bigg|_{r=r_{ph}} < 0 
		\ & \Leftrightarrow & \ 
		\mathcal{K}(r)\big|_{r=r_{ph}} < 0 
		\\
		\frac{d^{2}V_{\text{eff}}(r)}{dr^{2}} \bigg|_{r=r_{ph}} > 0 
		\ & \Leftrightarrow & \ 
		\mathcal{K}(r)\big|_{r=r_{ph}} > 0 
	\end{eqnarray}
\end{subequations}

The black hole shadows are closely connected with the unstable photon spheres. For a static and spherically symmetric black hole, an observer far away from the central black hole would observe the following angular radius of black hole shadow \cite{Perlick2022}
\begin{equation}
	\sin^{2}\alpha_{sh} = \frac{h(r_{ph})^{2}}{h(r_{\text{obs}})^{2}} = \frac{b^{2}_{\text{critical}}}{h(r_{\text{obs}})^{2}} 
\end{equation} 
where $r_{\text{obs}}$ is the radial coordinate of the observer, and the function $h(r)$ is defined by
\begin{equation}
	h(r) \equiv \sqrt{\frac{g_{\phi\phi}(r,\theta=\frac{\pi}{2})}{-g_{tt}(r)}} = \frac{r}{\sqrt{f(r)}}
\end{equation}
Meanwhile, the critical impact parameter $b_{\text{critical}}$ can also gives an absolute size / radius of the black hole shadow  \cite{Perlick2022}
\begin{equation}
	r_{sh}=b_{\text{critical}}=\frac{r_{ph}}{\sqrt{f(r_{ph})}}
\end{equation}
This is the analytical expression of black hole shadow radius for spherically symmetric black holes detected by an observer located at infinity. 

\subsection{Lens Equation and Einstein Ring in the Gravitational Lensing \label{sec:3d}}

\begin{figure}
	\includegraphics[width=1.00\textwidth]{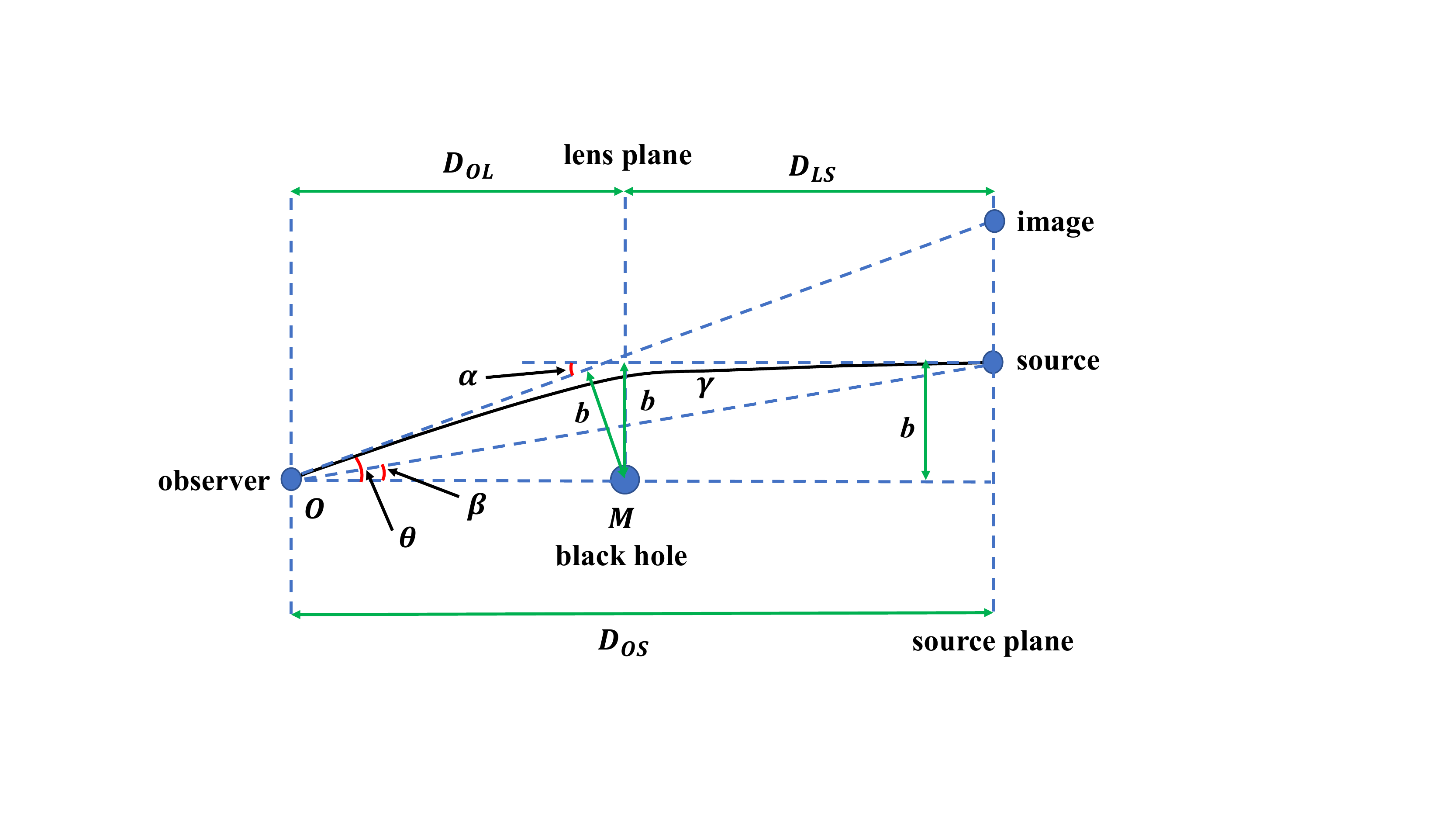}
	\caption{Schematically plot of the gravitational lensing. The photons emitted from a light source (luminous heavenly body) are lensed by the central supermassive black hole. In this figure, $D_{\text{OS}}$ is the distance between observer and source plane, $D_{\text{OL}}$ is the distance between observer and lens plane, $D_{\text{LS}}$ is the distance between lens plane and source plane. The central black hole is located in the middle of lens plane. The angle $\beta$ denotes the exact angular position of the light source with respect to axis $\text{OM}$, $\theta$ is the visual angular position of the lensed image seen by an observer, $\alpha$ is the gravitational deflection angle of light, and $b$ is the impact parameter.}
	\label{Weak gravitational lensing figure}
\end{figure}

In a galaxy, the light beams emitted from remote luminous light sources can be deflected or distorted due to the central supermassive black hole (which plays the role of a gravitational lens). Eventually, these luminous light sources may exhibit multiple images after light beams converging from the gravitational lens. The Einstein ring is one of such famous examples.

In gravitational lensing observations, the Einstein ring and other important physical observables are mostly constrained by the gravitational lens equation. The position of lensed objects and the Einstein ring can be calculated using the lens equation. There are a number of gravitational lens equations proposed in the past decades \cite{Virbhadra1998,Virbhadra2000,Frittelli1998,Dabrowski2000,Bozza2001,Perlick2004,Bozza2006,Bozza2008}. In the present work, we use a famous lens equation given by V. Bozza \cite{Bozza2008}
\begin{equation}
	D_{\text{OS}}\tan\beta = \frac{D_{\text{OL}}\sin\theta-D_{\text{LS}}\sin(\alpha-\theta)}{\cos(\alpha-\theta)} \label{lens equation}
\end{equation} 
Here, $D_{\text{OS}}$ is the distance between observer and source plane, $D_{\text{OL}}$ is the distance between observer and lens plane, $D_{\text{LS}}$ is the distance between lens plane and source plane. The angle $\beta$ in equation (\ref{lens equation}) denotes the exact angular position of the luminous light source, while the angle $\theta$ is the visual angular position of the lensed image seen by an observer faraway. From this lens equation, the images of the lensed objects can eventually be achieved. In the weak deflection limit, for distant sources and observers, we have the approximations $\tan\beta \approx \beta$, $\sin\theta \approx \theta$ and $\sin(\alpha-\theta) \approx \alpha-\theta$. Then the lens equation (\ref{lens equation}) reduces to the following form \cite{Bozza2001,Mollerach2002}
\begin{equation}
	\beta = \theta - \frac{D_{\text{LS}}}{D_{\text{OS}}} \cdot \alpha \label{lens equation reduced}
\end{equation} 	 
The relation $D_{\text{OS}} = D_{\text{OL}}+D_{\text{LS}}$ has been used to derive this equation. The angular radius of Einstein ring is calculated by taking $\beta = 0$.
\begin{equation}
	\theta_{\text{E}} = \frac{D_{\text{LS}}}{D_{\text{OS}}} \cdot \alpha 
	\label{Einstein ring}
\end{equation}
Further, in the weak deflection limit, the impact parameter $b$ satisfies
\begin{equation}
	b \approx D_{\text{OL}}\sin\theta_{\text{E}} \approx D_{\text{OL}}\theta_{\text{E}}
	\label{impact parameter approximation}
\end{equation}
Once the gravitational deflection angle $\alpha$ is known, the angular radius of Einstein ring $\theta_{\text{E}}$ can be solved combing the equation (\ref{Einstein ring}) and equation (\ref{impact parameter approximation}).

\section{Gravitational Deflection Angles of Light \label{sec:4}}

This section gives results and discussions on gravitational deflection angles of light for Schwarzschild and charged Reissner-Nordstr\"om (RN) black holes immersed in perfect fluid dark matter (PFDM). Subsection \ref{sec:4a} presents the analytical results obtained within the Gibbons and Werner approach by applying the Gauss-Bonnet theorem. Subsection \ref{sec:4b} displays the numerical results obtained using the conventional geodesic approach, in which the gravitational deflection angles are calculated by solving the trajectories of photon orbits. 

\subsection{Results from the Gibbons and Werner Approach (Gauss-Bonnet Theorem) \label{sec:4a}}

In the Gibbons and Werner approach, the gravitational deflection angle is calculated through a surface integral of Gaussian curvature in equation (\ref{Gauss-Bonnet gravitational deflection}). Here, we mainly focus on the analytical results of gravitational deflection angle for the charged Reissner-Nordstr\"om (RN) black hole immersed in PFDM. The analytical expressions for Schwarzschild black hole immersed in PFDM can be deduced from the charged black hole results by imposing $Q=0$.

For a charged RN black hole immersed in PFDM, the optical geometry restricted in equatorial plane ($\theta=\pi/2$) gives
\begin{eqnarray}
	dt^{2} & = & \tilde{g}_{ij}^{\text{OP}}dx^{i}dx^{j} 
	= \frac{1}{[f(r)]^2}dr^{2}+\frac{r^{2}}{f(r)}d\phi^{2} \nonumber
	\\
	& = & \frac{dr^{2}}{\big[1-\frac{2M}{r}+\frac{Q^{2}}{r^{2}}+\frac{\lambda_{\text{DM}}}{r}\cdot\ln\big(\frac{r}{|\lambda_{\text{DM}}|}\big)\big]^{2}}
	+\frac{r^{2}d\phi^{2}}{1-\frac{2M}{r}+\frac{Q^{2}}{r^{2}}+\frac{\lambda_{\text{DM}}}{r}\cdot\ln\big(\frac{r}{|\lambda_{\text{DM}}|}\big)} 
\end{eqnarray}
The Gaussian curvature in the equatorial plane of optical geometry can be expressed as \cite{Carmo1976}
\begin{eqnarray}
	\mathcal{K}(r) & = & -\frac{1}{\sqrt{\tilde{g}^{\text{OP}}}} 
	\bigg[ 
	\partial_{\phi} 
	\bigg(
	\frac{\partial_{\phi}\big(\sqrt{\tilde{g}^{\text{OP}}_{rr}}\big)}{\sqrt{\tilde{g}^{\text{OP}}_{\phi\phi}}}
	\bigg) 
	+\partial_{r} 
	\bigg(
	\frac{\partial_{r}\big(\sqrt{\tilde{g}^{\text{OP}}_{\phi\phi}}\big)}{\sqrt{\tilde{g}^{\text{OP}}_{rr}}}
	\bigg) 
	\bigg] \nonumber
	\\ 
	& = & \frac{1}{2} f(r) \cdot \frac{d^{2}f(r)}{dr^{2}}
	- \bigg[ \frac{1}{2} \cdot \frac{df(r)}{dr} \bigg]^{2} \nonumber \\
	& = & -\frac{4M+3\lambda_{\text{DM}}}{2r^{3}} 
	+\frac{12(M^{2}+Q^{2})+8M\lambda_{\text{DM}}-\lambda_{\text{DM}}^{2}}{4r^{4}} 
	-\frac{12MQ^{2}+Q^{2}\lambda_{\text{DM}}}{2r^{5}} +\frac{2Q^{4}}{r^{6}} \nonumber
	\\
	&   & +\bigg[
	\frac{\lambda_{\text{DM}}}{r^{3}}
	-\frac{\lambda_{\text{DM}}^{2}+3M\lambda_{\text{DM}}}{r^{4}} 
	+\frac{3Q^{2}\lambda_{\text{DM}}}{r^{5}}
	\bigg] \cdot  \ln\bigg(\frac{r}{|\lambda_{\text{DM}}|}\bigg) 
	+\frac{3\lambda_{\text{DM}}^{2}}{4r^{4}} \cdot \bigg[\ln\bigg(\frac{r}{|\lambda_{\text{DM}}|}\bigg)\bigg]^{2} \ \ \ \ \ 
	\label{Gaussian Curvature PFDM RN}        
\end{eqnarray}
and the surface area in the equatorial plane can be calculated through
\begin{eqnarray}
	dS & = &  \sqrt{\tilde{g}^{\text{OP}}} dr d\phi 
	=    \frac{r}{[f(r)]^{3/2}} dr d\phi   \nonumber
	\\
	& = & \bigg\{ 
	1 + \frac{3M}{r}  -\frac{3\lambda_{\text{DM}}}{2r}\cdot\ln\bigg(\frac{r}{|\lambda_{\text{DM}}|}\bigg) 
	- \frac{3Q^{2}}{2r^{2}}
	+\frac{15}{8} \bigg[\frac{2M}{r}-\frac{\lambda_{\text{DM}}}{r}\cdot\ln\bigg(\frac{r}{|\lambda_{\text{DM}}|}\bigg)\bigg]^2 \nonumber
	\\
	&   & +\text{higher order contributions}
	\bigg\}
	      \cdot rdrd\phi 
\end{eqnarray}
The higher order contributions in the surface area $dS$ include the $O\big(\frac{M^{3}}{r^{3}}\big)$,  $O\big(\frac{MQ^{2}}{r^{3}}\big)$, 
$O\big[\frac{M^{2}\lambda_{\text{DM}}}{r^{3}}\ln\big(\frac{r}{|\lambda_{\text{DM}}|}\big)\big]$,  $O\big[\frac{Q^{2}\lambda_{\text{DM}}}{r^{3}}\ln\big(\frac{r}{|\lambda_{\text{DM}}|}\big)\big]$, $O\big\{\frac{M\lambda_{\text{DM}}^{2}}{r^{3}} \big[\ln\big(\frac{r}{|\lambda_{\text{DM}}|}\big)\big]^{2}\big\}$ and $O\big\{\frac{\lambda_{\text{DM}}^{3}}{r^{3}} \big[\ln\big(\frac{r}{|\lambda_{\text{DM}}|}\big)\big]^{3}\big\}$.

In the optical geometry, the geodesic curvature $\kappa_{g}$ of the outer circular arc $C_{R}$ is calculated as follows \cite{Carmo1976}
\begin{eqnarray}
	\kappa_{g}(C_{R})
	& = & \frac{1}{2\sqrt{\tilde{g}^{\text{OP}}_{rr}}}
	\frac{\partial \ln \tilde{g}^{\text{OP}}_{\phi\phi}}{\partial r} \bigg|_{r=R} 
	=  \bigg[ 
	\frac{f(r)}{r} 
	-\frac{1}{2}\frac{\partial f(r)}{\partial r} 
	\bigg]_{r=R} \nonumber
	\\
	& = & \frac{1}{R}-\frac{6M+\lambda_{\text{DM}}}{2R^{2}}
	+\frac{3\lambda_{\text{DM}}}{2R^{2}}\cdot\ln\bigg(\frac{R}{|\lambda_{\text{DM}}|}\bigg)
	+\frac{2Q^{2}}{R^{3}} \label{geodesic curvature PFDM RN}
\end{eqnarray}
For the charged RN black hole immersed in PFDM, it can be easily verified that the corresponding optical geometry is an asymptotically flat space, because the Gaussian curvature and geodesic curvature in equations (\ref{Gaussian Curvature PFDM RN}) and (\ref{geodesic curvature PFDM RN}) both approach to zero in the $r \to \infty$ (or $R \to \infty$) limit. Based on the asymptotically flat property of the optical geometry, the contour integration of geodesic curvature $\kappa_{g}$ along the boundary $\partial D$ becomes
\begin{eqnarray}
	\int_{\partial D}\kappa_{g} dl
	& = & \lim_{R \to \infty} \int_{C_{R}}\kappa_{g}(C_{R}) \cdot dl 
	=  \lim_{R \to \infty} \int_{\phi_{\text{source}}}^{\phi_{\text{observer}}}\kappa_{g}(C_{R}) \cdot R d\phi \nonumber
	\\
	& \approx & \lim_{R \to \infty}
	\int_{0}^{\pi+\alpha}
	\bigg[
	\frac{1}{R}-\frac{6M+\lambda_{\text{DM}}}{2R^{2}} 
	+\frac{3\lambda_{\text{DM}}}{2R^{2}}\cdot\ln\bigg(\frac{R}{|\lambda_{\text{DM}}|}\bigg)
	+\frac{2Q^{2}}{R^{3}}
	\bigg] R d\phi \nonumber
	\\
	& = & \pi+\alpha \label{contour integral PFDM RN}
\end{eqnarray}
This result is consistent with the contour integral for asymptotically flat spacetimes summarized in equation (\ref{geodesic curvature contour integral}). In the present work, we also assume that the gravitational deflection angle is not large, which correspond to the weak deflection limit of the gravitational lensing. In these cases, the approximations $\phi_{\text{source}} \approx 0$ and $\phi_{\text{observer}} \approx \pi+\alpha$ can bu used for azimuthal angles.

Finally, the gravitational deflection angle of light for charged RN black hole immersed in PFDM halo is obtained from the integration of Gaussian curvature over the picked region $D$ in figure \ref{figure1}. As introduced in section \ref{sec:2}, it has been assumed that the black hole mass $M$, black hole charge $Q$ and dark matter parameter $\lambda_{\text{DM}}$ are in the same order of magnitude (namely $M \sim Q \sim \lambda_{\text{DM}}$). We can expand the gravitational deflection angle into series of small parameters $\epsilon \sim \frac{M}{b} \sim \frac{Q}{b} \sim \frac{\lambda_{\text{DM}}}{b}$ and $\eta \sim \epsilon \cdot \ln(\frac{1}{\epsilon}) \sim \frac{\lambda_{\text{DM}}}{b}\cdot \ln(\frac{b}{|\lambda_{\text{DM}}|})$.
\begin{eqnarray}
	\alpha & = & -\int_{D}\mathcal{K}(r)\cdot dS \nonumber
	\\
	& \approx & -\int_{0}^{\pi} d\phi 
	\int_{r(\gamma)}^{\infty} \mathcal{K}(r)\cdot\frac{r}{[f(r)]^{3/2}} dr   \nonumber
	\\
	& = & -\int_{0}^{\pi} d\phi
	\int_{r(\phi)}^{\infty} 
	\bigg\{ 
	-\frac{4M+3\lambda_{\text{DM}}}{2r^{2}} 
	+\frac{\lambda_{\text{DM}}}{r^{2}}\cdot\ln\bigg(\frac{r}{|\lambda_{\text{DM}}|}\bigg)
	-\frac{12(M^{2}-Q^{2})+10M\lambda_{\text{DM}}+\lambda_{\text{DM}}^{2}}{4r^3}  \nonumber
	\\
	&   &  \ \ \ \ \ \ \ \ \ \ \ \ \ \ \ \ \ \ \ \ \ 		
	+\frac{12M\lambda_{\text{DM}}+5\lambda_{\text{DM}}^{2}}{4r^{3}}\cdot\ln\bigg(\frac{r}{|\lambda_{\text{DM}}|}\bigg)
	-\frac{3\lambda_{\text{DM}}^{2}}{4r^{3}} \cdot \bigg[\ln\bigg(\frac{r}{|\lambda_{\text{DM}}|}\bigg)\bigg]^{2} 
	\bigg\} dr \nonumber
	\\
	&   &  \ \ \ \ \ \ \ \ \ \ \ \ \ \ \ \ \ \ \ \ \ 
	+\ \text{higher order contributions} \nonumber
	\\
	& = & \frac{4M-\lambda_{\text{DM}}}{b} 
	-\frac{2\lambda_{\text{DM}}}{b} \cdot \ln\bigg(\frac{b}{2|\lambda_{\text{DM}}|}\bigg) +\frac{15\pi M^{2}}{4b^{2}}-\frac{3\pi Q^{2}}{4b^{2}}+\frac{31\pi M\lambda_{\text{DM}}}{8b^{2}}+\frac{5\pi\lambda_{\text{DM}}^{2}}{32b^{2}} \cdot \bigg(\frac{\pi^{2}}{2}+1\bigg) \nonumber
	\\
	&   &   -\bigg[\frac{15\pi M}{4b} 
	+\frac{31\pi\lambda_{\text{DM}}}{16b} \bigg] \cdot \frac{\lambda_{\text{DM}}}{b} \ln\bigg(\frac{2b}{|\lambda_{\text{DM}}|}\bigg)
	+\frac{15\pi}{16}\frac{\lambda_{\text{DM}}^{2}}{b^{2}} \cdot
	\bigg[\ln\bigg(\frac{2b}{|\lambda_{\text{DM}}|}\bigg)\bigg]^{2} 
	+ O(\epsilon^{3}, \epsilon^{2}\eta, \epsilon\eta^{2}, \eta^{3}) \nonumber
	\\
	\label{gravitational deflection angle RN}
\end{eqnarray}
Here, the higher order contributions include the $\frac{M^{3}}{b^{3}}$, $\frac{MQ^{2}}{b^{3}}$, $\frac{M^{2}\lambda_{\text{DM}}}{b^{3}}$, $\frac{Q^{2}\lambda_{\text{DM}}}{b^{3}}$, $\frac{M\lambda_{\text{DM}}^{2}}{b^{3}}$, $\frac{\lambda_{\text{DM}}^{3}}{b^{3}}$ terms in order $O(\epsilon^{3})$, the $\frac{M^{2}\lambda_{\text{DM}}}{b^{3}}\ln(\frac{b}{|\lambda_{\text{DM}}|})$, $\frac{Q^{2}\lambda_{\text{DM}}}{b^{3}}\ln(\frac{b}{|\lambda_{\text{DM}}|})$, $\frac{M\lambda_{\text{DM}}^{2}}{b^{3}}\ln(\frac{b}{|\lambda_{\text{DM}}|})$, $\frac{\lambda_{\text{DM}}^{3}}{b^{3}}\ln(\frac{b}{|\lambda_{\text{DM}}|})$ terms in order $O(\epsilon^{2}\eta)$, the $\frac{M\lambda_{\text{DM}}^{2}}{b^{3}}\big[\ln(\frac{b}{|\lambda_{\text{DM}}|})\big]^{2}$, $\frac{\lambda_{\text{DM}}^{3}}{b^{3}}\big[\ln(\frac{b}{|\lambda_{\text{DM}}|})\big]^{2}$ terms in order $(\epsilon\eta^{2})$, and the $\frac{\lambda_{\text{DM}}^{3}}{b^{3}}\big[\ln(\frac{b}{|\lambda_{\text{DM}}|})\big]^{3}$ term in order $O(\eta^{3})$. The detailed expression of them would be very complicated. Notably, it is interesting to see that the higher order contributions from dark matter have the form $P(z_{1},z_{2},z_{3}) \cdot \frac{\lambda^{i}_{\text{DM}}}{b^{i}}\big[\ln(\frac{b}{|\lambda_{\text{DM}}|})\big]^{i}$ (with $P(z_{1},z_{2},z_{3})$ to be a polynomial function of $z_{1}=\frac{M}{b}$,$z_{2}=\frac{\lambda_{\text{DM}}}{b}$ and $z_{3}=\frac{Q^{2}}{b^{2}}$), such that each $\ln(\frac{b}{|\lambda_{\text{DM}}|})$ is always multiplied by a factor $\frac{\lambda_{\text{DM}}}{b}$.

In the above integration process, the photon orbit $r(\gamma)=r(\phi)$ in the lower limit of the integral should be known. In the Gibbons and Werner approach, the photon orbit $r(\phi)$ is usually replaced by an approximate solution. In this work, we choose the next-to-leading order approximate solution for photon orbit in the analytical calculation 
\begin{eqnarray}
	u(\phi) = \frac{1}{r(\phi)} 
	        & = & \frac{\sin\phi}{b} +\frac{M}{b^{2}}(1+\cos^{2}\phi) -\frac{\lambda_{\text{DM}}(1+\cos^{2}\phi)}{2b^{2}}\cdot\ln\bigg(\frac{b}{|\lambda_{\text{DM}}|\sin\phi}\bigg) 
	        \nonumber \\ 
	        &   & -\frac{\lambda_{\text{DM}}}{2b^{2}}\bigg[\cos^{2}\phi+2\cos\phi\cdot\ln\big(\tan\frac{\phi}{2}\big)\bigg] 
	        +O(\frac{\epsilon\eta}{b},\frac{\epsilon\eta}{b},\frac{\eta^{2}}{b}) 
	        \label{photon orbit approximation}
\end{eqnarray} 
where $b$ is impact parameter indicated in figure \ref{figure1} 
\footnote{The leading order approximation of photon orbit $u(\phi) = \frac{1}{r(\phi)} \approx \frac{\sin\phi}{b}$ can give the correct gravitational deflection angle only for $\frac{M}{b}$, $\frac{\lambda_{\text{DM}}}{b}$, $\frac{Q^{2}}{b^{2}}$ terms. If we are going to get all the correct next-to-leading order contributions in the gravitational deflection angle (especially the $\frac{M^{2}}{b^{2}}$, $\frac{M\lambda_{\text{DM}}}{b^{2}}$, $\frac{\lambda_{\text{DM}}^{2}}{b^{2}}$ terms in order $O(\epsilon^{2})$, $\frac{M\lambda_{\text{DM}}}{b^{2}}\ln\big(\frac{b}{|\lambda_{\text{DM}}|}\big)$,$\frac{\lambda_{\text{DM}}^{2}}{b^{2}}\ln\big(\frac{b}{|\lambda_{\text{DM}}|}\big)$ terms in order $O(\epsilon\eta)$, and $\frac{\lambda_{\text{DM}}^{2}}{b^{2}}\big[\ln\big(\frac{b}{\lambda_{|\text{DM}}|}\big)\big]^{2}$ term in order $O(\eta^{2})$), the next-to-leading order approximation for photon orbit $u(\phi)=\frac{1}{r(\phi)}$ should be included. The approximate solution in equation (\ref{photon orbit approximation}) can be solved by an iterative method.}. 
Notably, the gravitational deflection angle obtained in equation (\ref{gravitational deflection angle RN}) is consistent with the results given by F. Atamurotov \emph{et al.} and X. -J. Gao \emph{et al.} in references \cite{Atamurotov2022,YPHu2023}. Recently, F. Atamurotov gave the gravitational deflection angle for a rotating charged black hole immersed in PFDM \cite{Atamurotov2022}
\begin{equation}
	\alpha = 
	\frac{2M}{b} \bigg( 1+\frac{1}{v^{2}} \bigg)
	-\frac{\pi Q^{2}}{4b^{2}} \bigg( 1+\frac{2}{v^{2}} \bigg) +\frac{\lambda_{\text{DM}}}{v^{2}b} \bigg[ (1+v^{2})\ln2 -(1+v^{2})\ln\bigg(\frac{b}{|\lambda_{\text{DM}}|}\bigg) -v^{2} \bigg] 
	\pm \frac{4aM}{b^{2}v}   
\end{equation}
with $v=1$ for massless photons and $a=0$ for the non-rotating charged RN black hole immersed in PFDM. F. Atamurotov only gives the next-to-leading order contribution of gravitational deflection angle from the black hole charge (the $\frac{Q^{2}}{b^{2}}$ term). Our work also presents the next-to-leading order contributions from the black hole mass and dark matter halo. 
Furthermore, a second order gravitational deflection angle in the ``thin dark matter'' assumption $M \sim Q \gg \lambda_{\text{DM}}$ is calculated by X. -J. Gao \emph{et al.} \cite{YPHu2023}
\begin{equation}
	\alpha = \frac{2M}{b} +\frac{15\pi M^{2}}{4b^{2}} -\frac{3\pi Q^{2}}{4b^{2}}
	         - \frac{\lambda_{\text{DM}}}{b} \bigg[ 1-2\ln2+2\ln\bigg(\frac{b}{|\lambda_{\text{DM}}|}\bigg) \bigg]
\end{equation} 
which is exactly identical to our result in equation (\ref{gravitational deflection angle RN}) with $\frac{M\lambda_{\text{DM}}}{b^{2}}$, $\frac{\lambda_{\text{DM}}^{2}}{b^{2}}$, $\frac{M\lambda_{\text{DM}}}{b^{2}}\ln\big(\frac{b}{|\lambda_{\text{DM}}|}\big)$, $\frac{\lambda_{\text{DM}}^{2}}{b^{2}}\ln\big(\frac{b}{|\lambda_{\text{DM}}|}\big)$, $\frac{\lambda_{\text{DM}}^{2}}{b^{2}}\big[\ln\big(\frac{b}{|\lambda_{\text{DM}}|}\big)\big]^{2}$ omitted. These terms can be negligible in their ``thin dark matter'' assumption $M \sim Q \gg \lambda_{\text{DM}}$ \footnote{Using the ``thin dark matter'' approximation, X. -J. Gao \emph{et al.} proposed that $\frac{M}{b} \sim \frac{Q}{b} \sim \epsilon$ and $\frac{\lambda_{\text{DM}}}{b} \ll \epsilon$ (and Gao's choice of notation in reference \cite{YPHu2023} is $\frac{\lambda_{\text{DM}}}{b} \sim \epsilon^{2}$). So the $\frac{M\lambda_{\text{DM}}}{b^{2}} \ll \epsilon^{2}$ and $\frac{\lambda_{\text{DM}}^{2}}{b^{2}} \ll \epsilon^{2}$ terms in gravitational deflection angles can be neglected in the second order expansion. However, these contributions cannot be neglected in the second order expansion under our $M \sim \lambda_{\text{DM}} \sim Q$ assumption. A similar analysis shows that $\frac{M\lambda_{\text{DM}}}{b^{2}}\ln\big(\frac{b}{|\lambda_{\text{DM}}|}\big) $, $\frac{\lambda_{\text{DM}}^{2}}{b^{2}}\ln\big(\frac{b}{|\lambda_{\text{DM}}|}\big)$, $\frac{\lambda_{\text{DM}}^{2}}{b^{2}}\big[\ln\big(\frac{b}{|\lambda_{\text{DM}}|}\big)\big]^{2}$ terms can be neglected in their ``thin dark matter'' approximation (because $\frac{\lambda_{\text{DM}}}{b}\ln\big(\frac{b}{|\lambda_{\text{DM}}|}\big) \ll \epsilon \cdot \ln(\frac{1}{\epsilon}) \sim \eta$ satisfies in ``thin dark matter'' approximation).}. 
Therefore, our expression in equation (\ref{gravitational deflection angle RN}) could be viewed as an generalization of their results which goes beyond the ``thin dark matter'' approximation. Particularly, in the absence of dark matter halo (namely $\lambda_{\text{DM}}=0$), the gravitational deflection angle of light becomes $\alpha \approx \frac{4M}{b} +\frac{15\pi M^{2}-3\pi Q^{2}}{4b^{2}}$.

The Schwarzschild black hole immersed in PFDM is a special case of the charged black hole immersed in PFDM. Therefore, the gravitational deflection angle for Schwarzschild black hole immersed in PFDM can be obtained from equation (\ref{gravitational deflection angle RN}) by taking $Q=0$. This result is consistent with the gravitational deflection angle derived by S. Haroon \emph{et al.} in literature \cite{Haroon2019}. In 2019, S. Haroon  gave the leading order contribution of the gravitational deflection angle for a rotating black holes immersed in PFDM \cite{Haroon2019}
\begin{equation}
	\alpha = \frac{4M}{b} 
	-\frac{\lambda_{\text{DM}}}{b}\cdot
	\bigg[
	1-2\ln2+2\ln\bigg(\frac{b}{|\lambda_{\text{DM}}|}\bigg)
	\bigg]
	\pm \frac{4aM}{b^{2}}
\end{equation}
The $a=0$ corresponds to the non-rotating Schwarzschild black hole immersed in PFDM, which is perfectly agrees with our presented result in the leading order. 

\subsection{Results from the Conventional Geodesic Approach \label{sec:4b}}

In the conventional geodesic approach, the gravitational deflection angle is calculated by solving the trajectory of photon orbits near a black hole. Following the algorithm presented in subsection \ref{sec:3b}, the gravitational deflection angle of light can be obtained from the change of azimuthal angle $\Delta\phi$. In this work, we utilize the equation (\ref{gravitational deflection angle from null geodesic --- equivalent expression}) to numerically calculate the gravitational deflection angle when light source and observers are located at infinity.
\begin{equation}
	\alpha = 2\int_{0}^{1}\frac{dx}{\sqrt{F_{0}-F(x)}} -\pi 
	\label{gravitational deflection angle --- the conventional geodesic approach}
\end{equation}
Particularly, for the Schwarzschild black hole immersed in PFDM, the explicit expressions for $F_{0}$ and $F(x)$ are
\begin{subequations}
	\begin{eqnarray}
		F(x) & = & x^{2} - \frac{2M}{r_{0}} \cdot x^{3}
		+ \frac{\lambda_{\text{DM}}}{r_{0}} \cdot \ln\bigg(\frac{r_{0}}{|\lambda_{\text{DM}}|}\cdot\frac{1}{x}\bigg) \cdot x^{3} \ \ \ \ \ \ 
		\\
		F_{0} & = & 1 - \frac{2M}{r_{0}} + \frac{\lambda_{\text{DM}}}{r_{0}}\cdot \ln\bigg(\frac{r_{0}}{|\lambda_{\text{DM}}|}\bigg)
	\end{eqnarray}
\end{subequations}
For the charged RN black hole immersed in PFDM, the explicit expressions for $F_{0}$ and $F(x)$ are
\begin{subequations}
	\begin{eqnarray}
		F(x) & = & x^{2} - \frac{2M}{r_{0}} \cdot x^{3}
		+ \frac{Q^{2}}{r_{0}^2} \cdot x^{4} 
		+ \frac{\lambda_{\text{DM}}}{r_{0}} \cdot \ln\bigg(\frac{r_{0}}{|\lambda_{\text{DM}}|}\cdot\frac{1}{x}\bigg) \cdot x^{3} 
		\\
		F_{0} & = & 1 - \frac{2M}{r_{0}} + \frac{Q^{2}}{r_{0}^{2}} + \frac{\lambda_{\text{DM}}}{r_{0}}\cdot \ln\bigg(\frac{r_{0}}{|\lambda_{\text{DM}}|}\bigg)
	\end{eqnarray}
\end{subequations}
Here $r_{0}$ is the minimal distance to central supermassive black hole in the photon orbit, and the variable $x$ is defined as $x=\frac{r_{0}}{r}$. Particularly, in the absence of dark matter, the function $F(x)$ becomes a polynomial function, which makes the gravitational deflection angle expressed by a hyper-elliptic integral
\begin{equation}
	F=\int_{x_{1}}^{x_{2}}\frac{dx}{\sqrt{a-P_{n}(x)}}
\end{equation}
Here, $a$ is a constant, $P_{n}(x)$ is a polynomial function of $x$ in n-th order. The existence of dark matter halo makes the function $F(x)$ more complicated.

In the above integration process, the closet distance $r_{0}$ to central supermassive black hole in the photon orbit can be solved from the impact parameter $b$ via the equation
\begin{equation}
	b^{2}=\frac{L^{2}}{E^{2}}=\frac{r_{0}^{2}}{f(r_{0})}
\end{equation}
Furthermore, when the closet distance $r_{0}$ approaches to the radius of unstable photon sphere (namely $r_{0} \to r_{ph}$), the gravitational deflection angle of light would become divergent \cite{Bozza2008}, and the corresponding critical impact parameter is the shadow radius detected by an observer at infinity ($r_{sh}=b_{\text{critical}}$). The gravitational deflection in this case is extremely strong.

\begin{figure*}
	\includegraphics[width=0.55\textwidth]{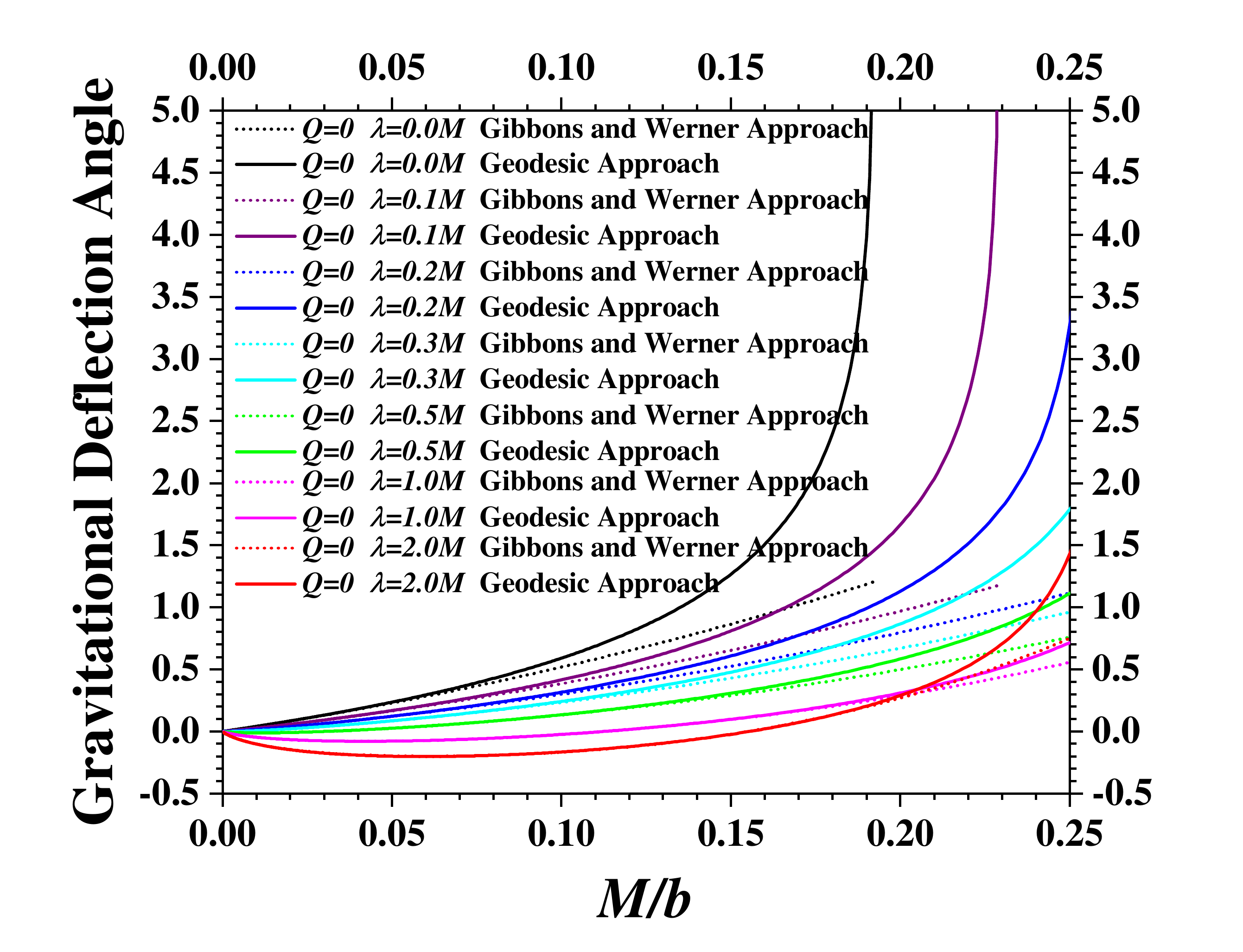}
	\includegraphics[width=0.55\textwidth]{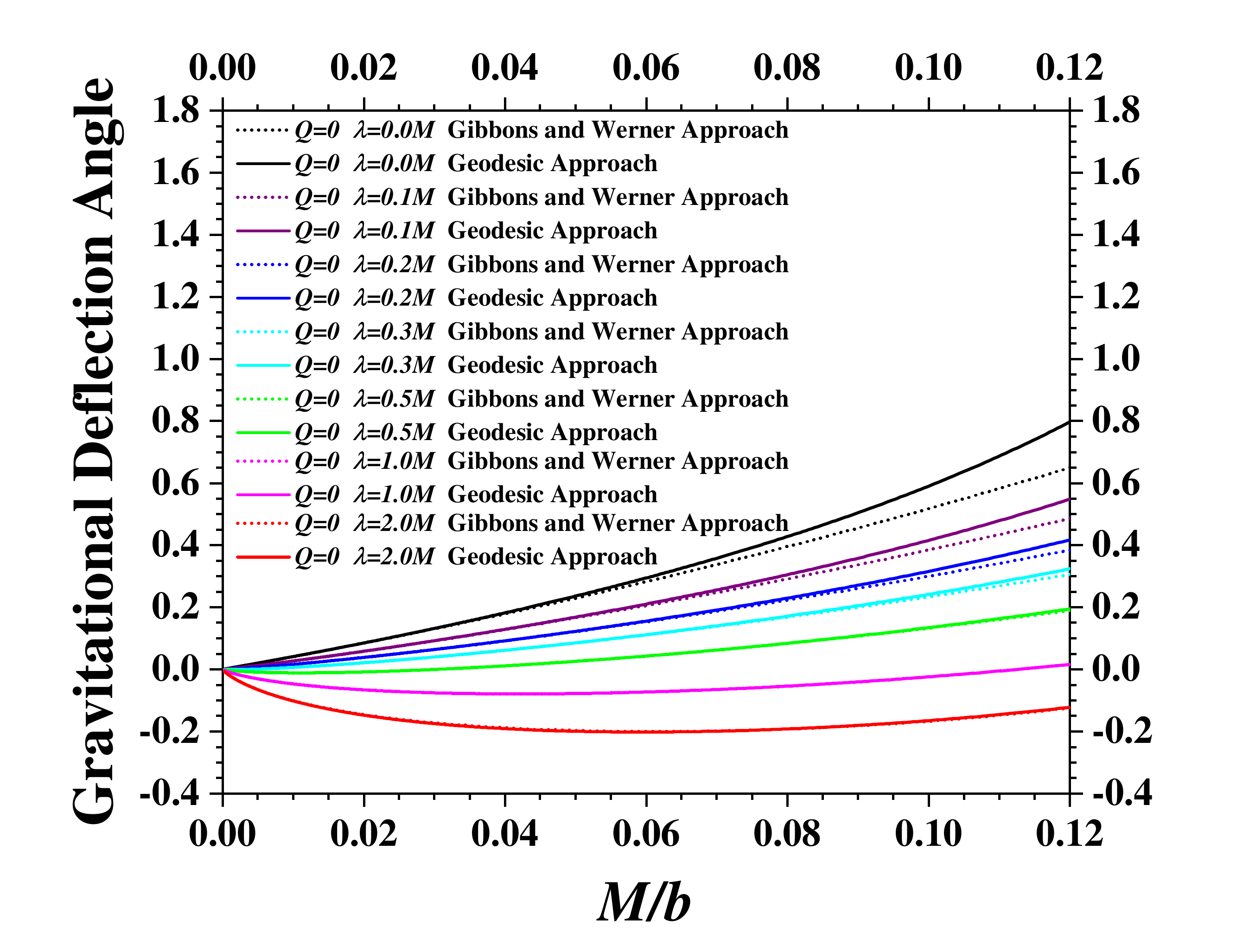}
	\caption{The gravitational deflection angles for Schwarzschild black hole immersed in PFDM. This figure present the numerical results calculated from the Gibbons and Werner approach using equation (\ref{gravitational deflection angle RN}) (with black hole charge $Q=0$) and results from the conventional geodesic approach using equation (\ref{gravitational deflection angle --- the conventional geodesic approach}). Here, the dark matter parameters are selected to be $\lambda_{\text{DM}}=0$, $\lambda_{\text{DM}}=0.1M$, $\lambda_{\text{DM}}=0.2M$, $\lambda_{\text{DM}}=0.3M$, $\lambda_{\text{DM}}=0.5M$, $\lambda_{\text{DM}}=1.0M$, $\lambda_{\text{DM}}=2.0M$ to highlight the effects of dark matter halo on the gravitational deflection angle. The right panel shows the amplified region where $\frac{M}{b}$ is not very large.}
	\label{figure3_ab}
\end{figure*}

\begin{figure*}
	\includegraphics[width=0.55\textwidth]{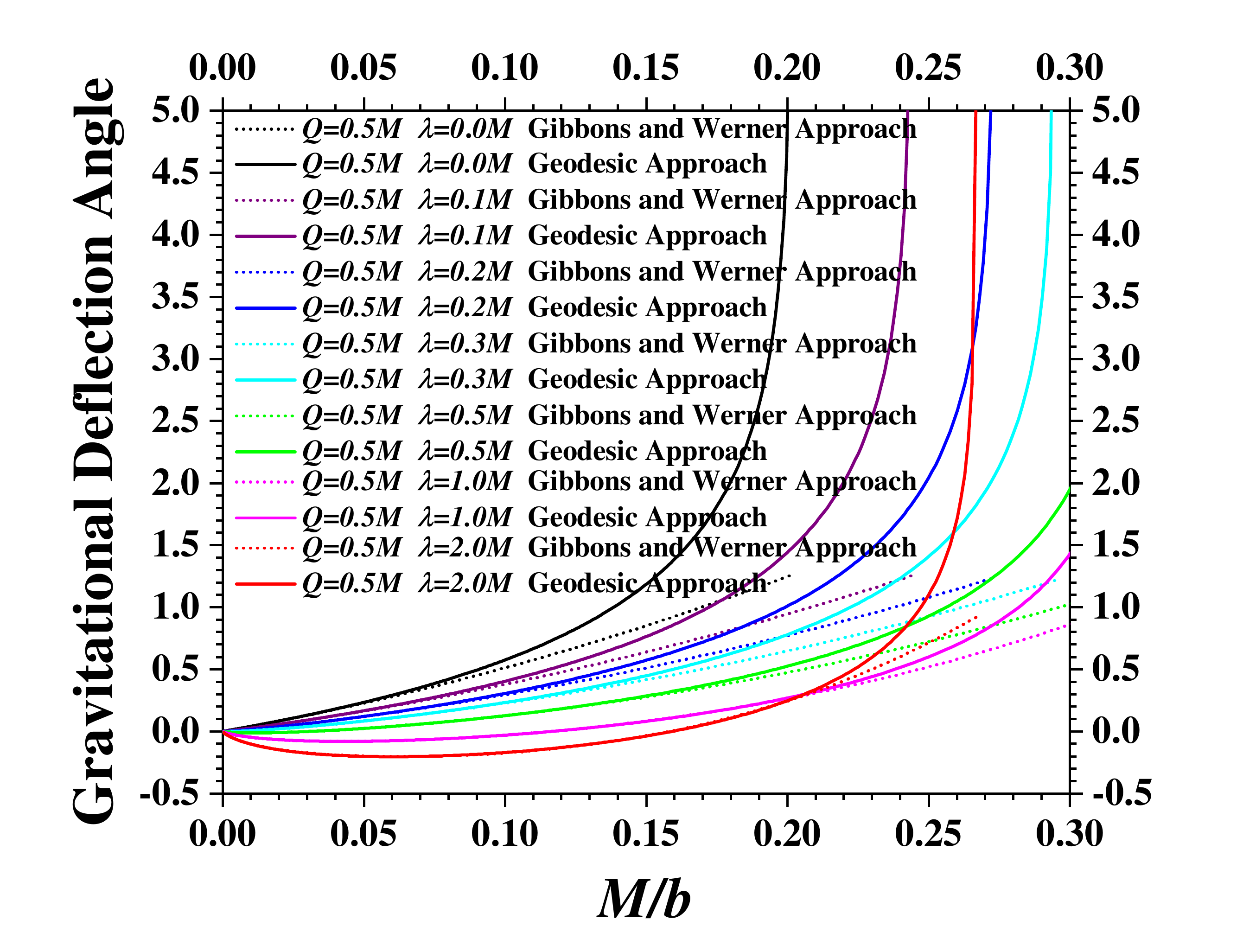}
	\includegraphics[width=0.55\textwidth]{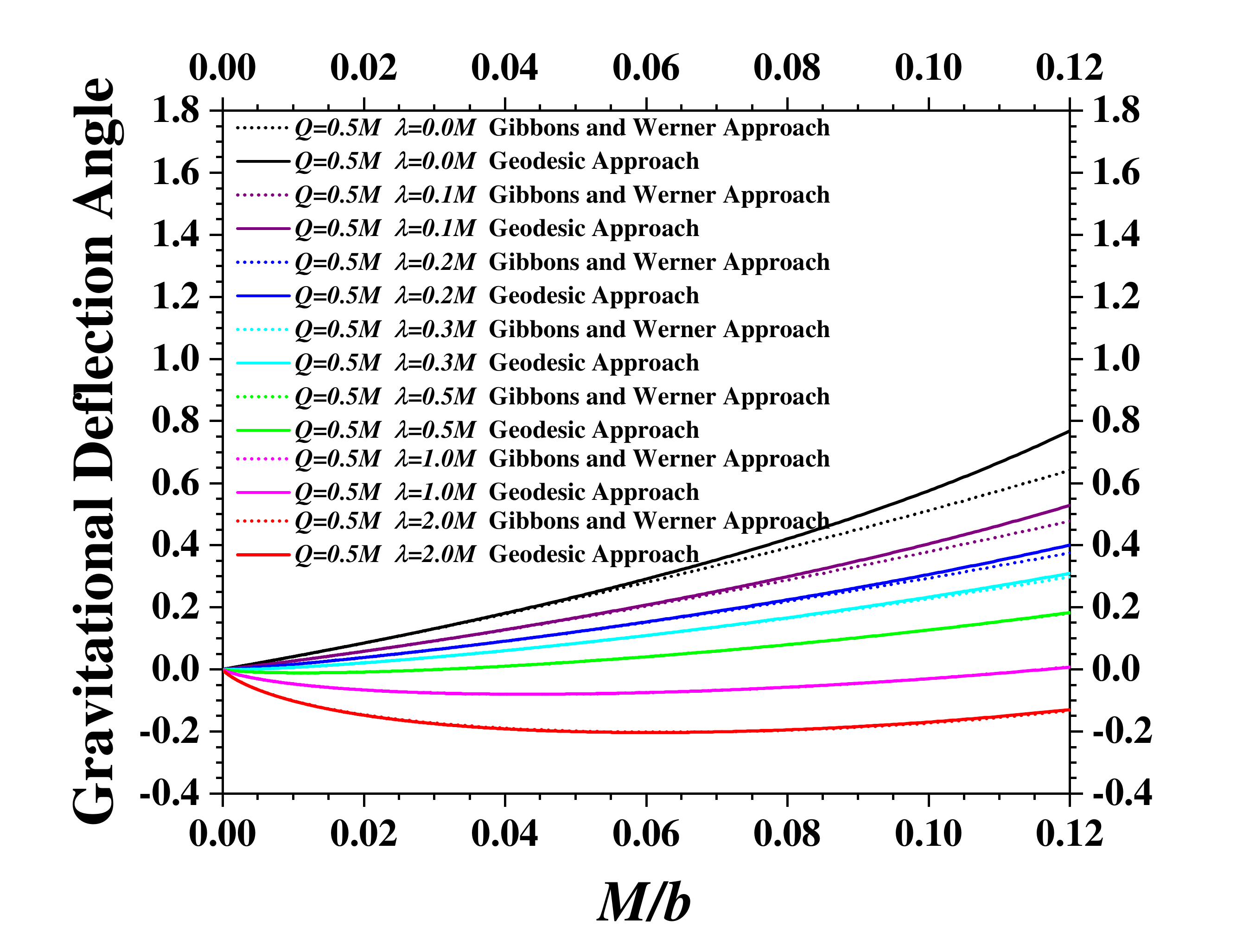}
	\caption{The gravitational deflection angles for charged RN black hole immersed in PFDM. This figure present the numerical results on gravitational deflection angles calculated from the Gibbons and Werner approach using equation (\ref{gravitational deflection angle RN}) and results from the conventional geodesic approach using equation (\ref{gravitational deflection angle --- the conventional geodesic approach}). In this figure, the black hole charge is chosen to be $Q=0.5M$, and the dark matter parameters are selected as $\lambda_{\text{DM}}=0$, $\lambda_{\text{DM}}=0.1M$, $\lambda_{\text{DM}}=0.2M$, $\lambda=0.3M$, $\lambda_{\text{DM}}=0.5M$, $\lambda_{\text{DM}}=1.0M$, $\lambda_{\text{DM}}=2.0M$ to highlight the effects of dark matter halo on gravitational deflection angle. The right panel shows the amplified region where $\frac{M}{b}$ is not very large. }
	\label{figure3_cd}
\end{figure*}

\begin{figure*}
	\includegraphics[width=0.55\textwidth]{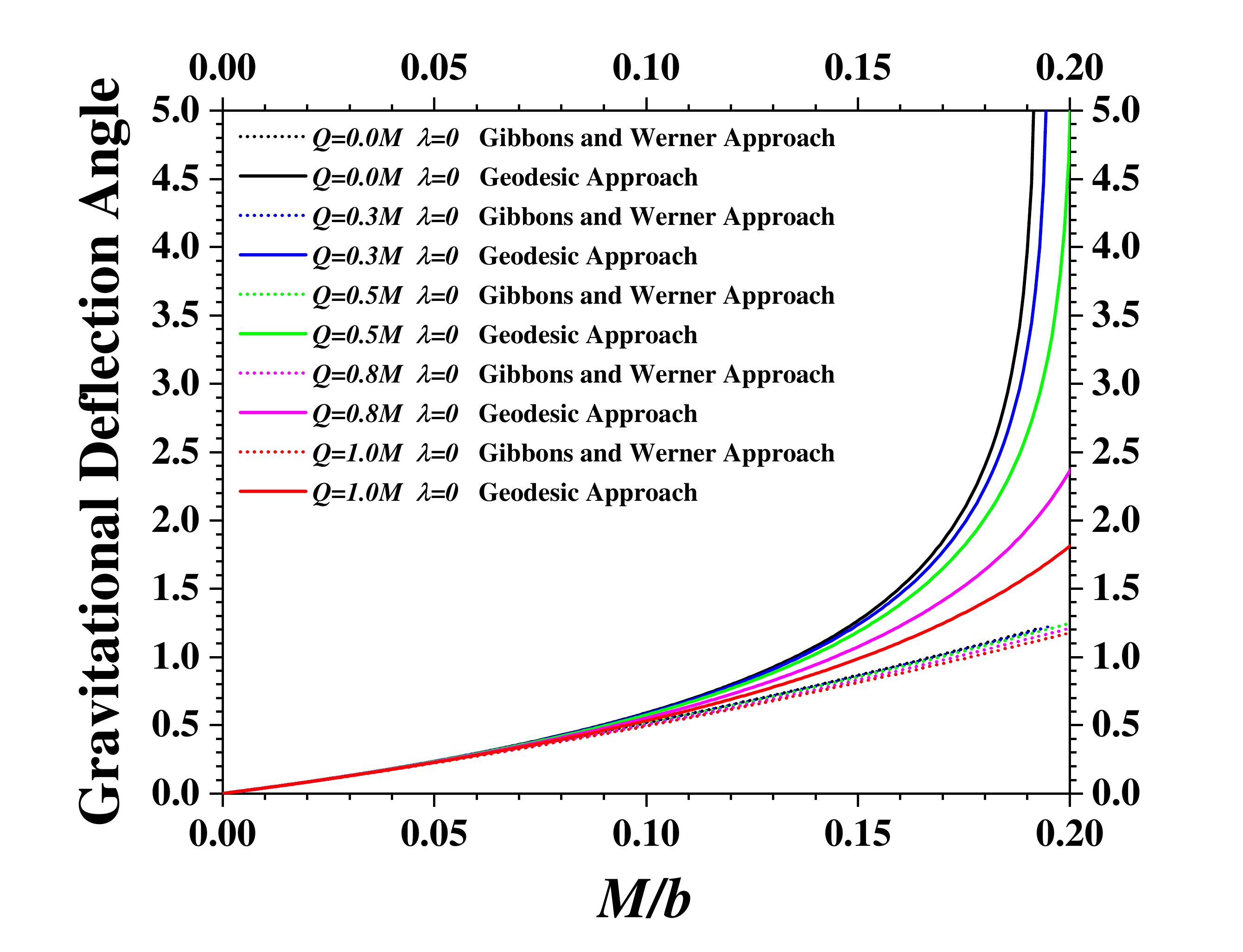}
	\includegraphics[width=0.55\textwidth]{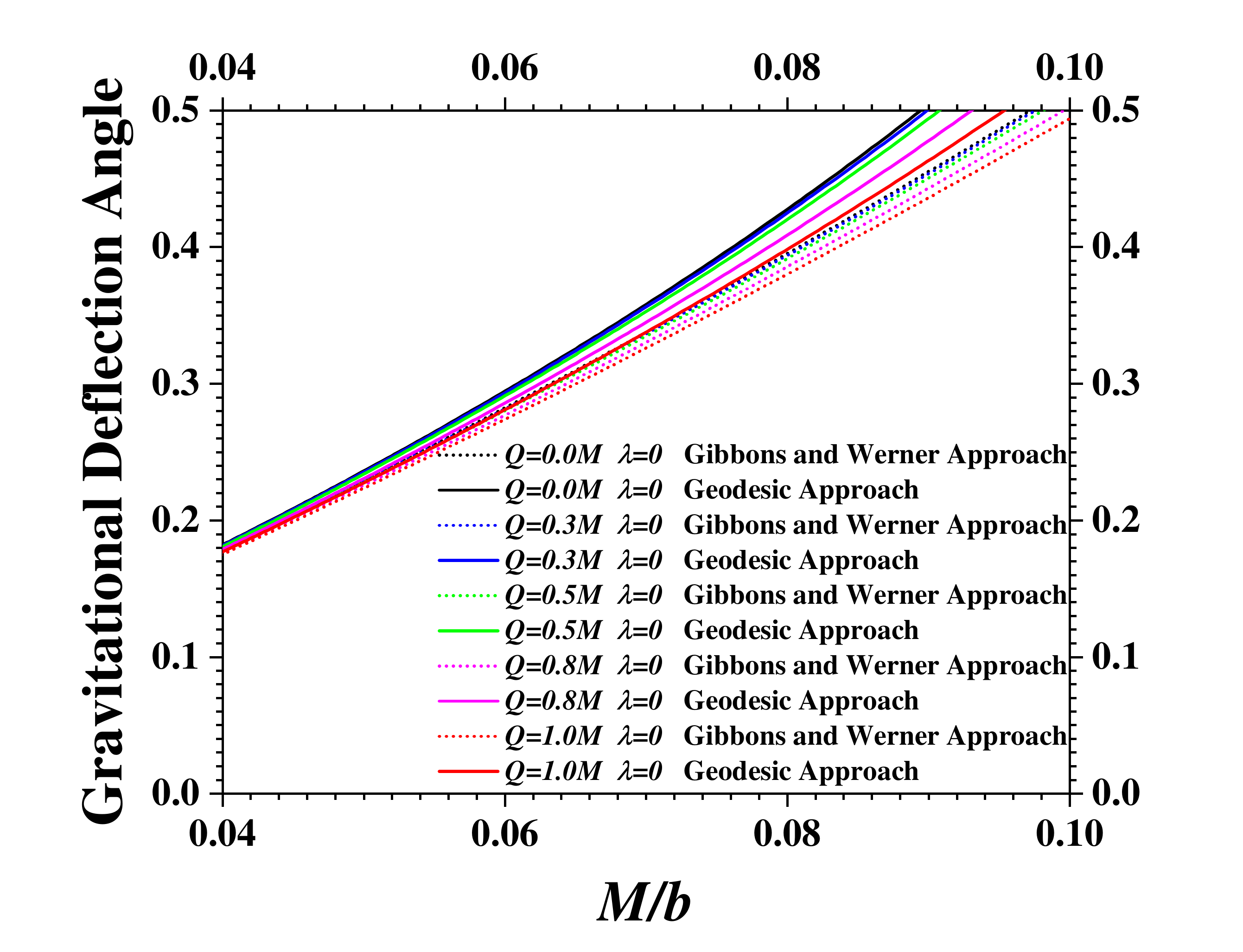}
	\caption{The gravitational deflection angles for charged RN black hole immersed in PFDM. This figure present the numerical results calculated from the Gibbons and Werner approach in equation (\ref{gravitational deflection angle RN}) and results from the conventional geodesic approach in equation (\ref{gravitational deflection angle --- the conventional geodesic approach}). To highlight the effects coming from black hole charge $Q$ on the gravitational deflection angle, in this figure, the dark matter parameter is selected as $\lambda_{\text{DM}}=0$, and the black hole charges are chosen to be $Q=0$, $Q=0.3M$, $Q=0.5M$, $Q=0.8M$, $Q=1.0M$ respectively. The right panel shows the amplified region such that the numerical results obtained from the Gibbons and Werner approach for different black hole charges can be effectively distinguished.}
	\label{figure3_ef}
\end{figure*}

\begin{figure*}
	\includegraphics[width=0.55\textwidth]{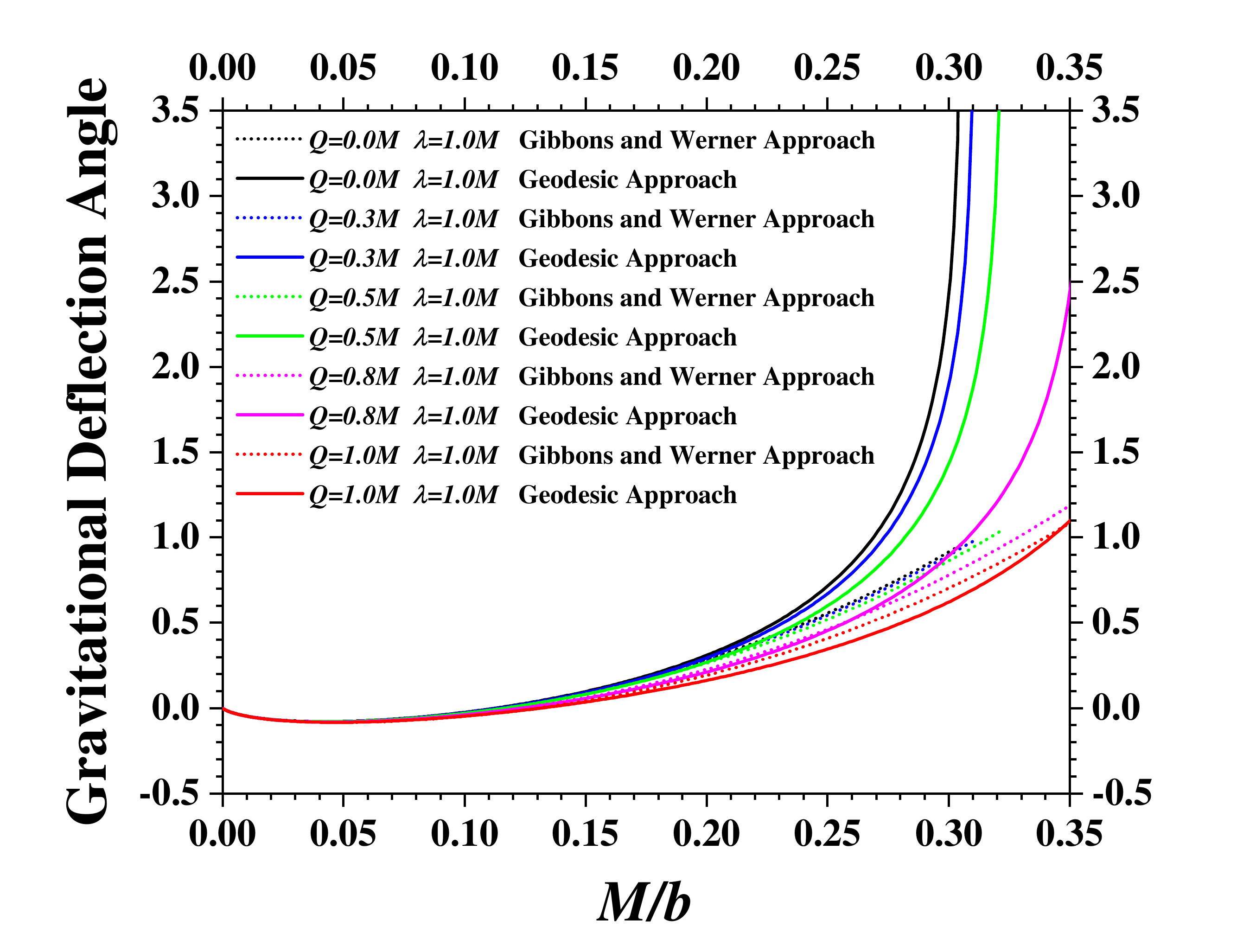}
	\includegraphics[width=0.55\textwidth]{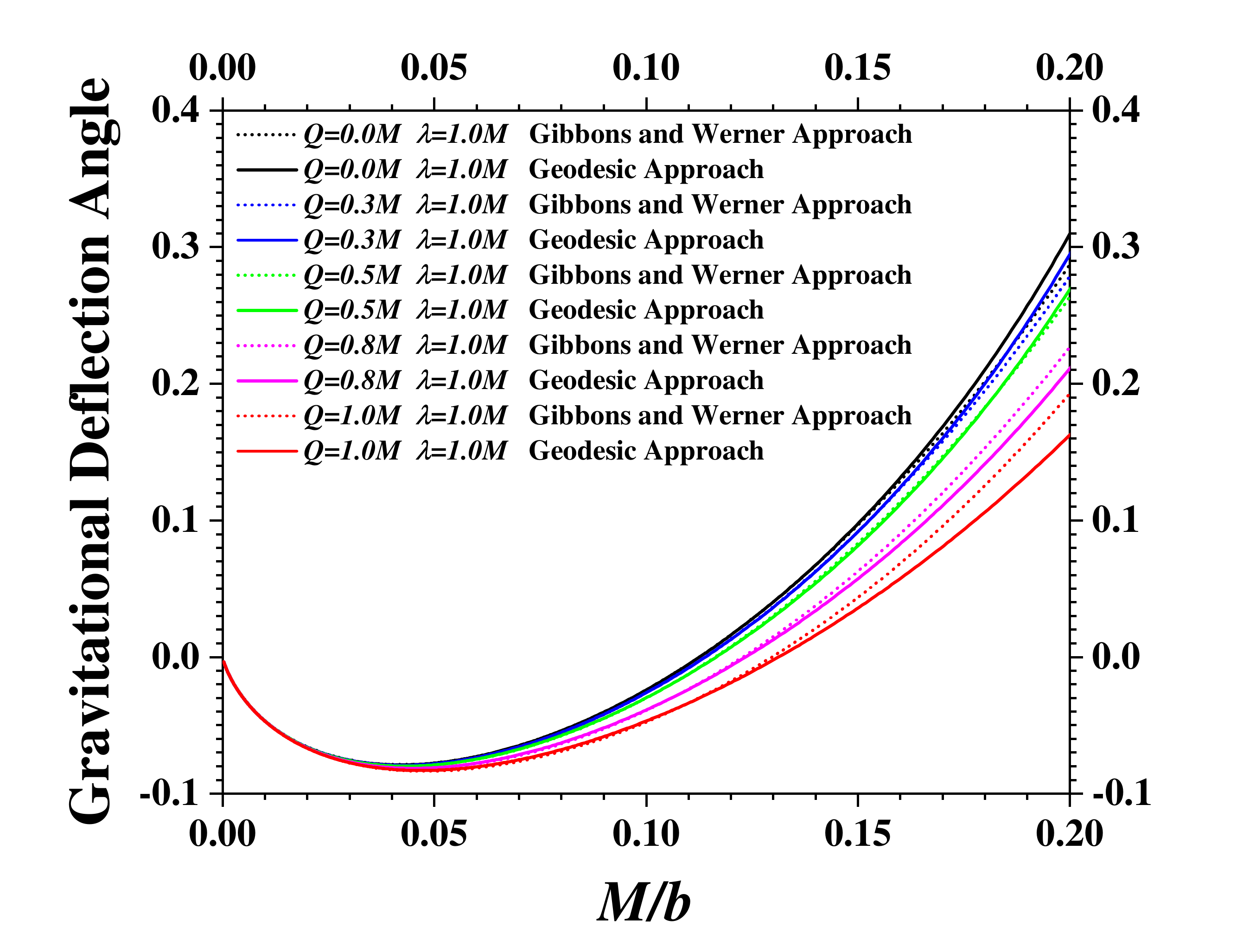}
	\caption{The gravitational deflection angles for charged RN black hole immersed in PFDM. This figure present the numerical results calculated from the Gibbons and Werner approach using equation (\ref{gravitational deflection angle RN}) and results from the conventional geodesic approach using equation (\ref{gravitational deflection angle --- the conventional geodesic approach}). To emphasize the effects coming from black hole charge $Q$ on the gravitational deflection angle, in this figure, we select the dark matter parameter $\lambda_{\text{DM}}=1.0M$, and the black hole charges are chosen to be $Q=0$, $Q=0.3M$, $Q=0.5M$, $Q=0.8M$ and $Q=1.0M$. The right panel shows the amplified region where $\frac{M}{b}$ is not very large.}
	\label{figure3_gh}
\end{figure*}

The numerical results on gravitational deflection angles of light for the Schwarzschild and charged RN black holes immersed in PFDM are plotted in figures \ref{figure3_ab}-\ref{figure3_gh}. The results calculated from the Gibbons and Werner approach and results from the conventional geodesic approach are presented for comparisons. In the weak deflection cases where $\frac{M}{b}$, $\frac{\lambda_{\text{DM}}}{b}$ are very small, the gravitational deflection angles calculated from the Gibbons and Werner approach and those from the conventional geodesic approach agree with each other. However, in the strong deflection cases where $\frac{M}{b}$, $\frac{\lambda_{\text{DM}}}{b}$ are large, the gravitational deflection angles obtained from two approaches may exhibit notable differences. There are two reasons lead to this discrepancy. Firstly, the gravitational deflection angle given in equation  (\ref{gravitational deflection angle RN}) (which is obtained from the Gibbons and Werner approach using the Gauss-Bonnet theorem) is a second-order expansion of $\frac{M}{b}$ (or $\frac{\lambda_{\text{DM}}}{b}$). Higher-order contributions, which are non-negligible when $\frac{M}{b}$, $\frac{\lambda_{\text{DM}}}{b}$ are larger, are not included in equation  (\ref{gravitational deflection angle RN}). Secondly, in the strong deflection limit, the photon orbits could be dramatically distorted, which makes them completely different with the slightly bent photon orbits illustrated in figure \ref{figure1}. In these cases, the approximation $\phi_{\text{observer}} \approx \pi + \alpha \approx \pi$ used in the integration of Gaussian curvature no longer valid. From the comparisons in figures \ref{figure3_ab}-\ref{figure3_gh}, it can be concluded that the Gibbons and Werner approach using Gauss-Bonnet theorem could underestimate the gravitational deflection angle in the strong deflection limit.

Particularly, the figure \ref{figure3_ab} and figure \ref{figure3_cd} emphasize the effects of dark matter halo on the gravitational deflection angle, while figure \ref{figure3_ef} and figure \ref{figure3_gh} highlight the influences coming from black hole charge $Q$. From these figures, we can easily find that the black hole charge has negative contributions on gravitational deflection angles. For the same dark matter parameter $\lambda_{\text{DM}}$, black holes with larger electric charge would result in smaller deflection angles. Furthermore, in the weak deflection cases where $\frac{M}{b}$ is not large, the dark matter also has negative contributions on gravitational deflection angle, similar to the influences of black hole charge. From the right panels of figure \ref{figure3_ab} and figure \ref{figure3_cd}, for a fixed black hole charge, black holes with larger dark matter parameter $\lambda_{\text{DM}}$ eventually get smaller gravitational deflection angles (in the region $\frac{M}{b}$ is not very large). This is mostly caused by the leading order contribution $-\frac{\lambda_{\text{DM}}}{b}-\frac{2\lambda_{\text{DM}}}{b}\cdot\ln\big(\frac{b}{|2\lambda_{\text{DM}}|}\big)$ in the analytical expression of the gravitational deflection angle in equation (\ref{gravitational deflection angle RN}).

\begin{figure}
	\includegraphics[width=0.7\textwidth]{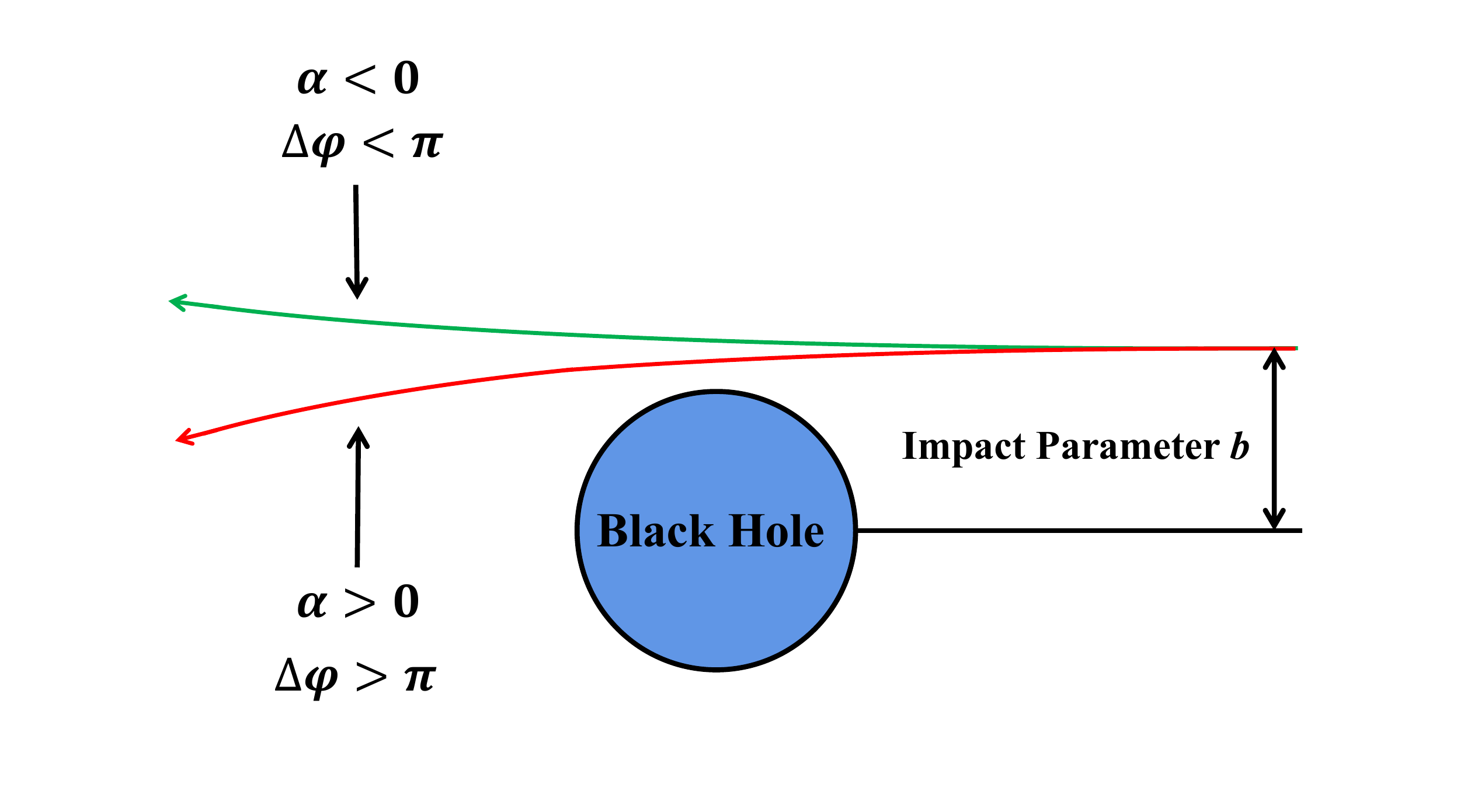}
	\caption{The positive and negative deflection angles of light in the gravitational lensing observations.}
	\label{positive and negative}
\end{figure}

In addition, from figures \ref{figure3_ab}-\ref{figure3_gh}, we also discover an interesting phenomenon. When the dark matter parameter is sufficiently large ($\lambda_{\text{DM}}>0.5M$), the gravitational deflection angle could become negative when $\frac{M}{b}$ is not very large. The positive and negative gravitational deflection angles can be interpreted in the following way. The positive deflection angle ($\alpha>0$) implies that the change of azimuthal angle in a photon orbit is greater than 180$^{o}$ ($\Delta\phi>\pi$), and the gravitational field produced by central black hole as well as the PFDM halo eventually give ``attractive effects'' on light beams. On the other hand, the negative gravitational deflection angle ($\alpha<0$) implies that the change of azimuthal angle in a photon orbit is less than 180$^{o}$ ($\Delta\phi<\pi$), and the gravitational field produced by central black hole as well as the PFDM halo eventually give ``repulsive effects'' on light beams. The positive and negative gravitational deflection angles are illustrated in figure \ref{positive and negative}.

\section{Photon Sphere and Black Hole Shadow \label{sec:5}}

In this section, we present the numerical results on photon spheres and black hole shadow radius for the Schwarzschild and charged RN black holes immersed in PFDM. These results can be obtained through the conventional approach by solving the local extreme points of effective potentials, or through an geometric approach developed by Qiao \emph{et al} using the Gaussian curvature and geodesic curvature in the equatorial plane of optical geometry. 

\begin{figure*}
	\includegraphics[width=0.55\textwidth]{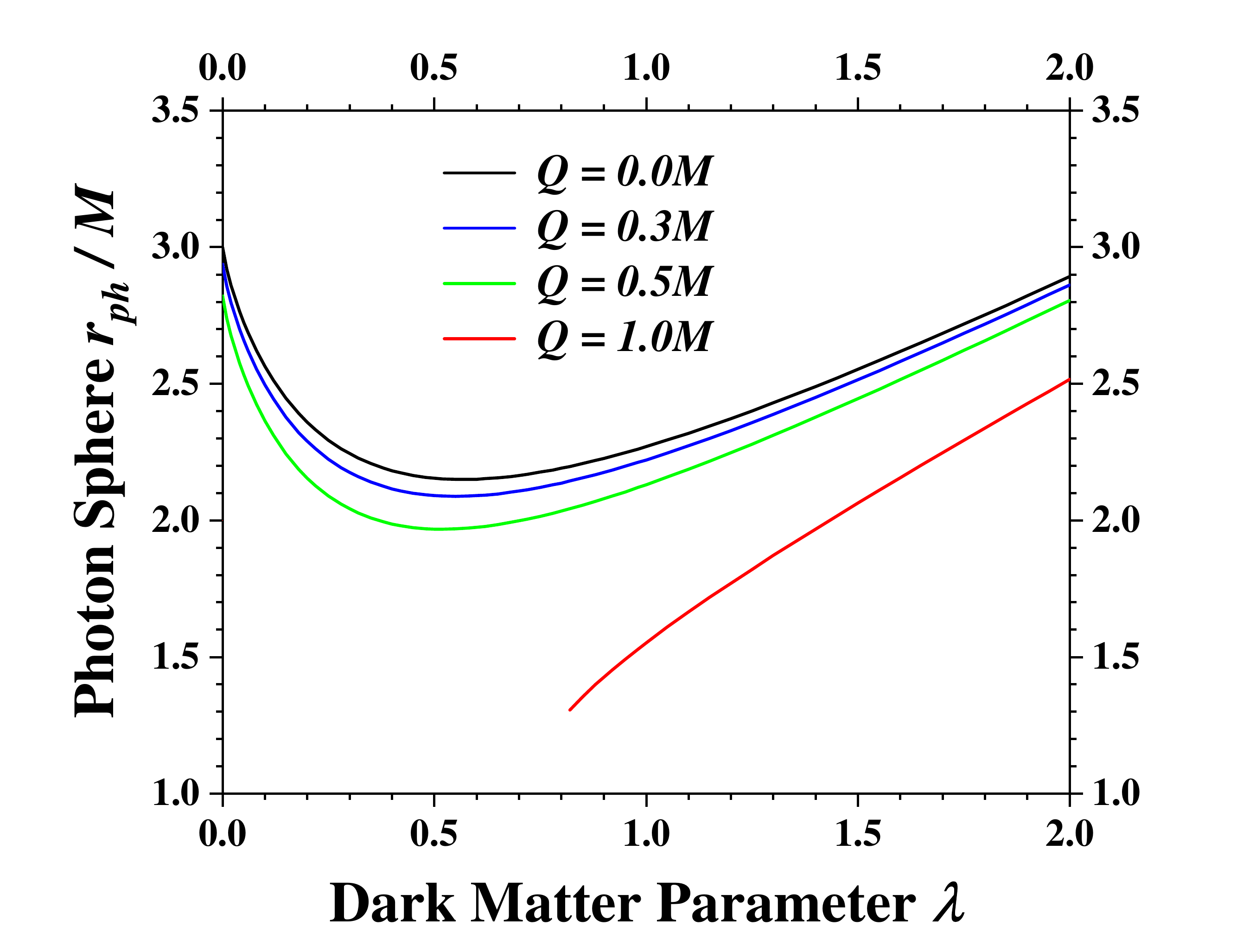}
	\includegraphics[width=0.55\textwidth]{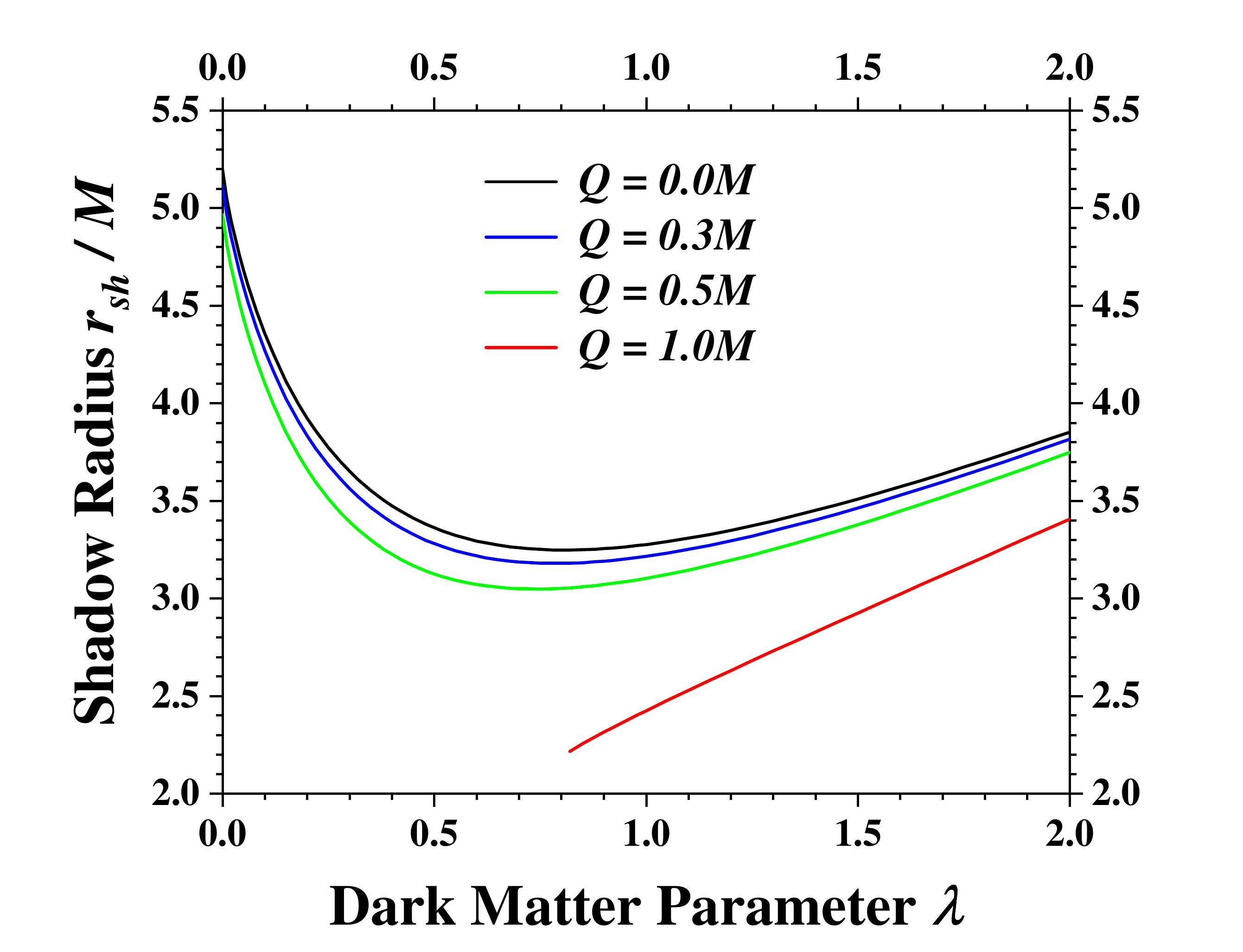}
	\caption{The unstable photon spheres and black hole shadows for the Schwarzschild and charged RN black holes immersed in PFDM. In this figure, we choose the black hole charge to be $Q=0$, $Q=0.2M$, $Q=0.5M$, $Q=M$ to show the unstable photon sphere and black hole shadow as the dark matter parameter varies from $\lambda_{\text{DM}}=0$ to $\lambda_{\text{DM}}=2M$. The left panel presents the unstable photon sphere radius $r_{ph}$, and the right panel shows the black hole shadow radius $r_{sh}=b_{\text{critical}}$ detected by an observer located at infinity. The $Q=0$ case represents the Schwarzschild black hole immersed in PFDM, and the $Q=M$ case corresponds to the extreme charged RN black hole immersed in PFDM.}
	\label{figure4}
\end{figure*}

\begin{figure*}
	\includegraphics[width=0.55\textwidth]{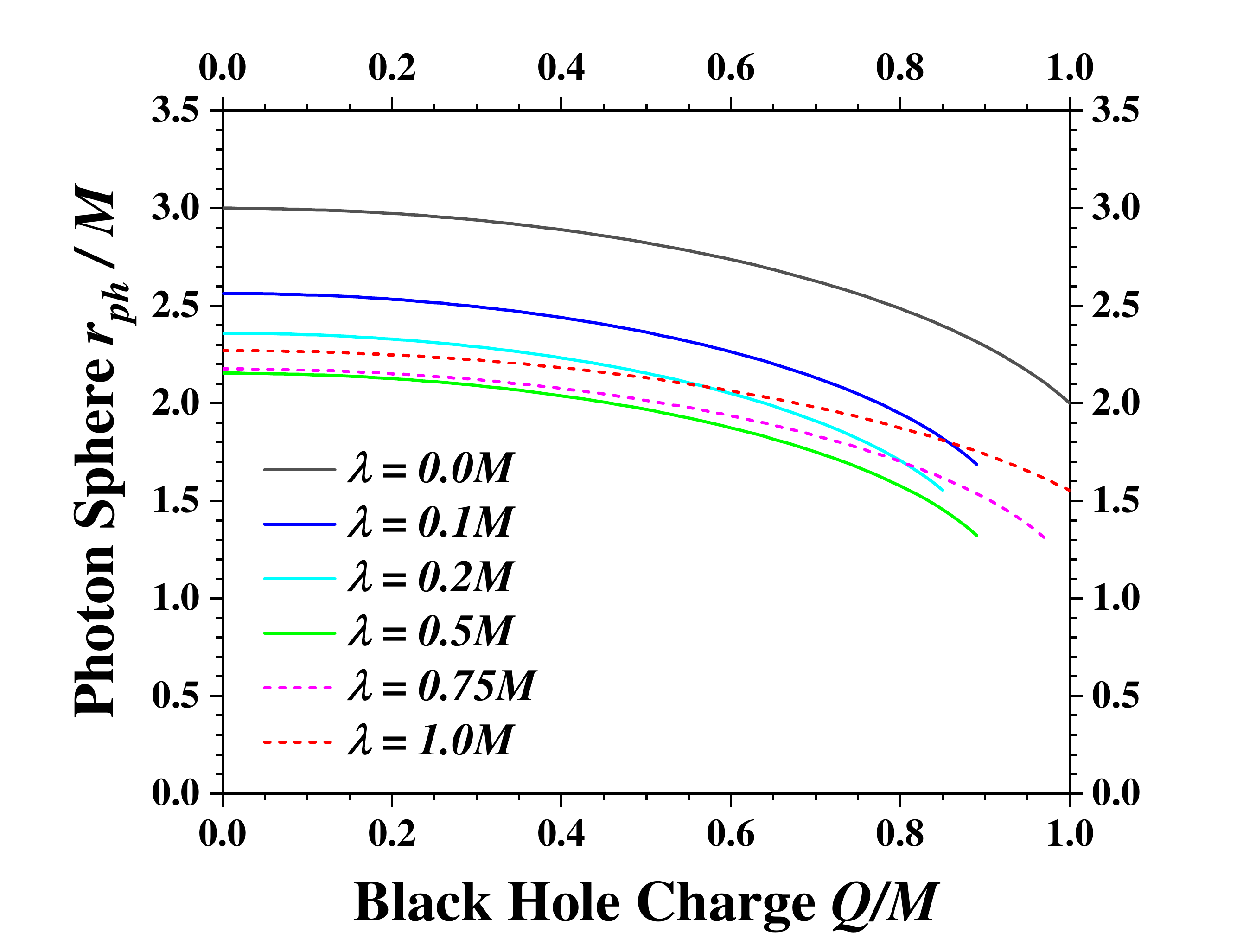}
	\includegraphics[width=0.55\textwidth]{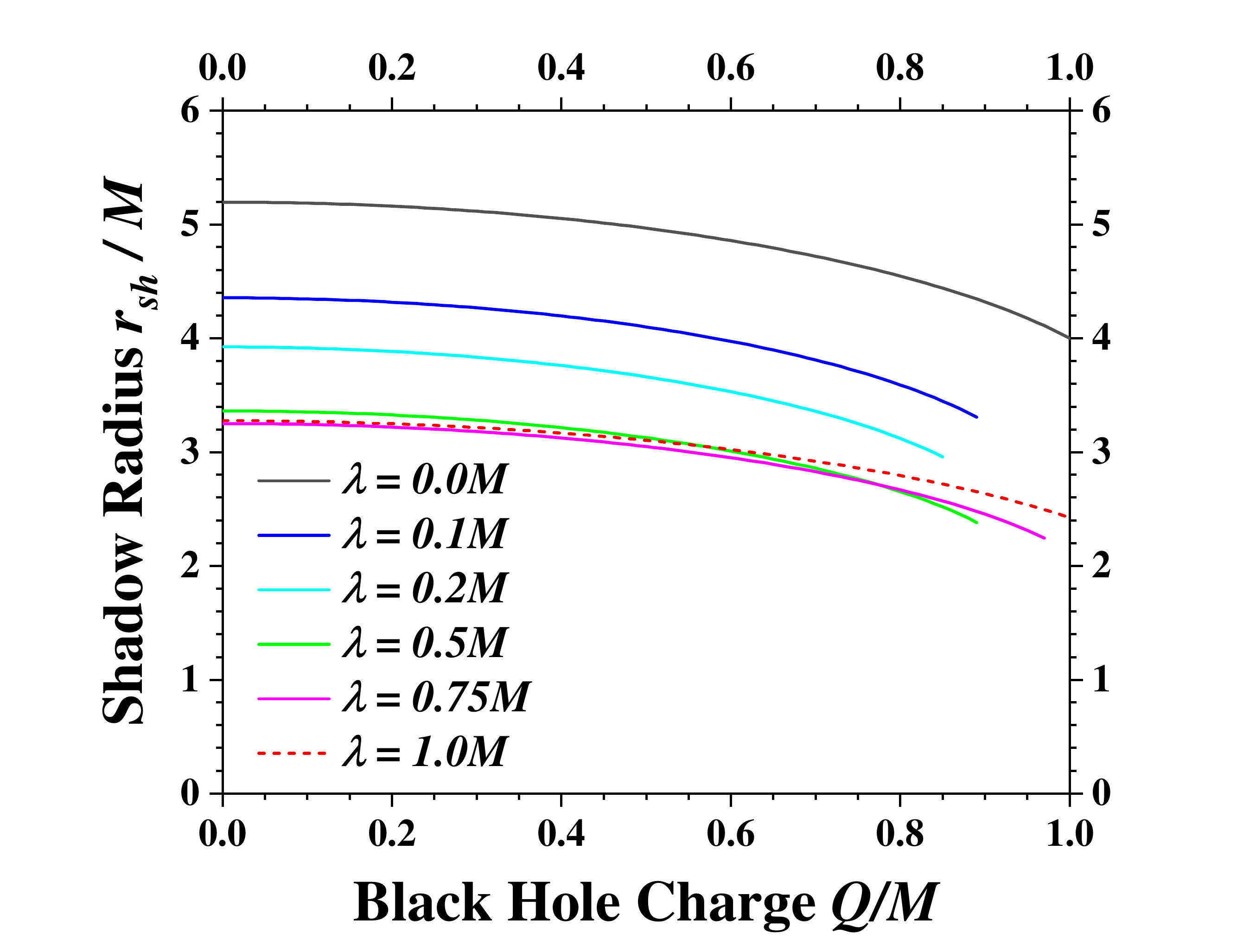}
	\caption{The unstable photon spheres and black hole shadows for the charged RN black hole immersed in PFDM. In this figure, we choose the dark matter parameter to be $\lambda_{\text{DM}}=0$, $\lambda_{\text{DM}}=0.1M$, $\lambda_{\text{DM}}=0.2M$, $\lambda_{\text{DM}}=0.5M$, $\lambda_{\text{DM}}=0.75M$, $\lambda_{\text{DM}}=M$ to plot the unstable photon sphere and black hole shadow as the black hole charge varies from $Q=0$ to $Q=M$. The left panel presents the unstable photon sphere radius $r_{ph}$, and the right panel displays the black hole shadow radius $r_{sh}=b_{\text{critical}}$ detected by an observer located at infinity. 
	The case $\lambda_{\text{DM}}=0$ corresponds to no dark matter medium, which recovers the standard RN black hole in general relativity. }
	\label{figure5}
\end{figure*}

The figure \ref{figure4} and figure \ref{figure5} plot the photon sphere $r_{ph}$ and black hole shadow radius $r_{sh}=b_{\text{critical}}$ when black hole charge $Q$ and dark matter parameter $\lambda_{\text{DM}}$ are varying. From these figures, it is clearly observed that the PFDM can greatly change the photon sphere and black hole shadow for central black holes. From figure \ref{figure4}, for a fixed black hole charge $Q$, when the dark matter parameter varies from $\lambda_{\text{DM}}=0$ to $\lambda_{\text{DM}}=2M$, both the photon sphere and black hole shadow first diminish and then get enlarged. The turning points are $\lambda_{\text{DM}} \approx 0.5M$ for photon spheres and $\lambda_{\text{DM}} \approx 0.75M$ for black hole shadows. When the dark matter parameter $\lambda_{\text{DM}}$ exceeds these tuning points, the photon sphere and black hole shadow begin to increase. 

In the figure \ref{figure5}, it is indicated that the black hole charge has the effects to reduce the photon sphere and black hole shadow. For the same dark matter parameter $\lambda_{\text{DM}}$, black hole with larger electric charge has smaller photon sphere and shadow radius. Another interesting phenomenon is that, in the extreme charged RN black hole case ($M=Q$), the black hole shadow disappears for some dark matter parameters ($0<\lambda_{\text{DM}}<M$), making the cut-off of the red curve in figure \ref{figure4}, as well as the cut-off of the green, cyan, blue, magenta curves in figure \ref{figure5} near the $Q\approx M$ region. This is mainly caused by the naked singularity in spacetime. In the extreme black hole case, the charged RN black hole immersed in PFDM could become horizonless and produce a naked singularity for dark matter parameter $0<\lambda_{\text{DM}}<M$. Furthermore, in the naked singularity cases (the extreme black hole $M=Q$ with dark matter parameter $0<\lambda_{\text{DM}}<M$), the photon spheres could have either no solution or multiple solutions. In other words, the photon spheres could disappear for some dark matter parameter values (in range $0<\lambda_{\text{DM}}<M$), while multiple photon spheres would emerge for other dark matter parameter values (also in range $0<\lambda_{\text{DM}}<M$)
\footnote{Actually, when the charged RN black hole approaches to the extreme black hole (namely $Q \sim M$), the naked singularities and multiple photon spheres have already appeared when dark matter parameter $0<\lambda_{\text{DM}}<M$. For simplicity, we only pick the extreme black hole case to plot in figure \ref{figure4}.}. 
However, we do not plot these multiple photon spheres in figure \ref{figure4}, since they are not always connected with black hole shadows 
\footnote{Note that only the unstable photon spheres in black hole spacetime are connect with black hole shadows, the stable photon spheres are not connected with black hole shadows.}. 
This unexpected behavior (the emergence of naked singularity and multiple photon spheres) may probably be caused by some unknown mechanisms or black hole phase transitions, and they deserve more intensive studies in the future.

\section{Gravitational Lens Equation and Einstein Ring \label{sec:6}}

In this section, we provide discussions on a significantly valuable observable in the gravitational lensing observations —— the Einstein ring. In the gravitational lensing, the Einstein ring angle is usually solved from the gravitational lens equation.

In this section, we mainly focus on the weak deflection limit, and the reduced gravitational lens equation in equation (\ref{lens equation reduced}) is used in the numerical calculations. Solving the reduced lens equation with $\beta=0$, the Einstein ring angle $\theta_{\text{E}}$ for the Schwarzschild black hole immersed in PFDM can be calculated as
\begin{eqnarray}
	\theta_{\text{E}} & = & \frac{D_{\text{LS}}}{D_{\text{OS}}} \cdot \alpha \nonumber
	\\
	& \approx & \frac{D_{\text{LS}}}{D_{\text{OS}}} \cdot
	\bigg\{
	\frac{4M-\lambda_{\text{DM}}}{b} 
	-\frac{2\lambda_{\text{DM}}}{b} \cdot \ln\bigg(\frac{b}{2|\lambda_{\text{DM}}|}\bigg) 
	+\frac{15\pi M^{2} - 3\pi Q^{2}}{4b^{2}} 
	+\frac{31\pi M\lambda_{\text{DM}}}{8b^{2}}  \nonumber
	\\
	&   &   +\frac{5\pi\lambda_{\text{DM}}^{2}}{32b^{2}} \cdot \bigg(\frac{\pi^{2}}{2}+1\bigg)
	-\bigg[\frac{15\pi M\lambda_{\text{DM}}}{4b^{2}} 
	+\frac{31\pi\lambda_{\text{DM}}^{2}}{16b^{2}} \bigg] \cdot \ln\bigg(\frac{2b}{|\lambda_{\text{DM}}|}\bigg)
	+\frac{15\pi\lambda_{\text{DM}}^{2}}{16b^{2}} \cdot
	\bigg[\ln\bigg(\frac{2b}{|\lambda_{\text{DM}}|}\bigg)\bigg]^{2}
	\bigg\} \nonumber
	\\
	 \label{Einstein Angle1}
\end{eqnarray}
Further, in the weak deflection limit, the Einstein ring angle $\theta_{E}$ is usually very small, and the impact parameter $b$ in gravitational lensing satisfies
\begin{equation}
	b \approx D_{\text{OL}}\sin\theta_{\text{E}} \approx D_{\text{OL}}\theta_{\text{E}} \label{Einstein Angle2}
\end{equation}
Finally, the Einstein ring angle of a lensed luminous object can be calculated by solving the equations (\ref{Einstein Angle1}) and (\ref{Einstein Angle2}).

\begin{figure}
	\includegraphics[width=0.7\textwidth]{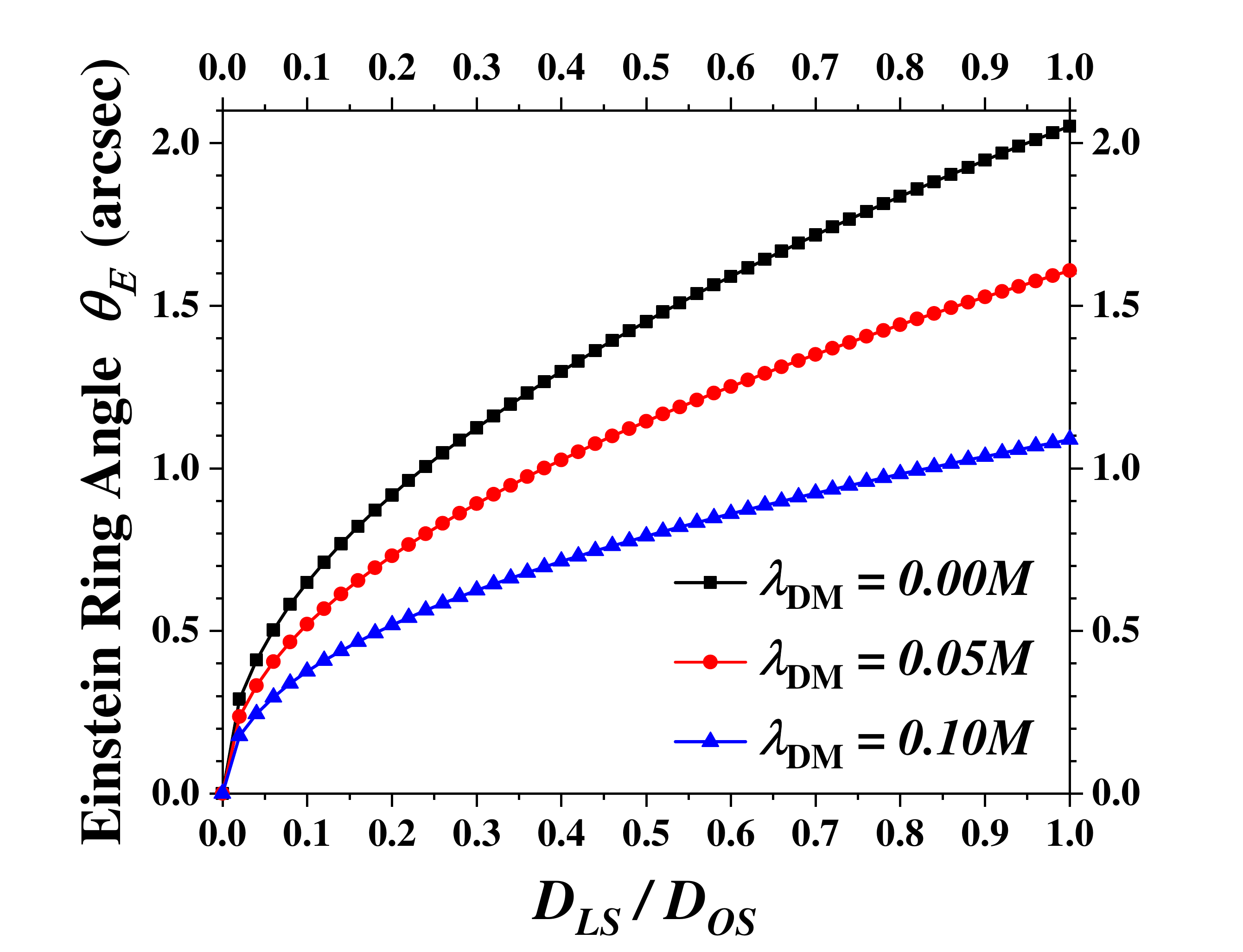}
	\caption{The Einstein ring angle of the Schwarzschild black hole immersed in PFDM. In this figure, the dark matter parameters are chosen to be $\lambda_{\text{DM}}=0$, $\lambda_{\text{DM}}=0.05M$ and $\lambda_{\text{DM}}=0.1M$. The horizontal axis labels the ratio of $D_{\text{LS}}/ D_{\text{OS}}$, and the vertical axis labels the Einstein ring angle $\theta_{\text{E}}$ in unit of arc-second. Here, the black hole mass is $M=4.3 \times 10^{6} M_{\odot}$, and the distance between the observer and lensed plane is $D_{\text{OL}}=D_{\text{LS}}-D_{\text{OS}}=8.33$ kpc.}
	\label{figure Einstein Angle}
\end{figure}

The figure \ref{figure Einstein Angle} presents the Einstein ring angle for Schwarzschild black hole immersed in PFDM, focusing the dark matter effects on central supermassive black holes. In this figure, the black hole mass is selected to be $M = 4.3 \times 10^{6}M_{\odot}$, and the distance between observer and lensed plane is $D_{\text{OL}}=8.33$ kpc (which corresponds to the mass and distance of Sgr A* in our galaxy) 
\footnote{Since there are no evidences that the supermassive black hole of Sgr A* in our galaxy has a non-negligible electric charged, we do not assign a nonzero black hole charge $Q$ in the numerical calculations of Einstein ring angle.}.  
The dark matter parameters are chosen as $\lambda_{\text{DM}}=0$, $\lambda_{\text{DM}}=0.05M$ and $\lambda_{\text{DM}}=0.1M$ respectively. The numerical results on Einstein ring angle are plotted with different ratios $D_{\text{LS}} / D_{\text{OS}}$. From this figure, it is clearly indicated that, in the presence of dark mater halo, the Einstein ring becomes smaller than that of conventional Schwarzschild black hole. For the same dark matter parameter, the size of Einstein ring increases when $D_{\text{LS}} / D_{\text{OS}}$ is larger. In order to observe notable Einstein ring images, the distance $D_{\text{LS}}$ should be as large as possible. Furthermore, when dark matter parameter $\lambda_{\text{DM}}$ is very large, the Einstein ring angle $\theta_{\text{E}}$ in equations (\ref{Einstein Angle1}) and (\ref{Einstein Angle2}) has no solutions. Therefore, the Einstein ring angles in the larger dark matter parameter cases are not plotted in figure \ref{figure Einstein Angle}.

\section{Summary and Conclusions \label{sec:7}}

In this work, we study the gravitational lensing of black holes immersed in a dark matter halo. We choose two typical black holes --- the Schwarzschild black hole and the charged Reissner-Nordstr\"om (RN) black hole --- to carry out the investigation. To analyze the influences from dark matter in a simple way, we assume that the dark matter is made up of perfect fluids with the energy momentum tensor satisfies $T_{\mu\nu}=(\rho+p)u_{\mu}u_{\nu}+p g_{\mu\nu}$. The spacetime metric for the Schwarzschild and charged RN black holes immersed in perfect fluid dark matter (PFDM) are parameterized by the black hole mass $M$, black hole charge $Q$ and dark matter parameter $\lambda_{\text{DM}}$. These two black hole examples could reflect many universal properties of the more complex black holes surrounded by dark matter halos, and they shall give us insights on the behaviors of the supermassive black hole in the galaxy center with a dark matter halo.

In the present work, the gravitational deflection angle of light, photon sphere, black hole shadow and Einstein ring angle are calculated and discussed under the assumption $M \sim \lambda_{\text{DM}} \sim Q$. Particularly, in the calculation of gravitational deflection angles, two approaches have been used. One is the Gibbons and Werner approach, in which the gravitational deflection angle is obtained using the Gauss-Bonnet theorem. The other one is the conventional geodesic approach, in which the gravitational deflection angle is obtained by solving the trajectories of null geodesics. Specifically, a second order analytical expansion of gravitational deflection angle for Schwarzschild and charged black holes immersed in PFDM is obtained using the Gibbons and Werner approach in the weak deflection limit, and the full gravitational deflection angle (including all order perturbation contributions applicable to both weak and strong deflection limits) is also calculated numerically within the conventional geodesic approach. In the weak deflection limit where $\frac{M}{b}$, $\frac{\lambda_{\text{DM}}}{b}$ are small, the numerical results from these two approaches agree with each other. However, in the strong deflection limit where $\frac{M}{b}$, $\frac{\lambda_{\text{DM}}}{b}$ are not small, there are non non-negligible discrepancies between results obtained using two different approaches. In the strong deflection limit, some approximations used in the integration of Gaussian curvature are no longer satisfied. Detailed comparisons show that the Gibbons and Werner approach using the Gauss-Bonnet theorem could underestimate the gravitational deflection angle in these cases.  

Under the assumption $M \sim \lambda_{\text{DM}} \sim Q$, the dark matter halo has a great effect on the gravitational lensing of central supermassive black holes. For larger dark matter parameters $\lambda_{\text{DM}}$, the Schwarzschild and charged RN black holes immersed in PFDM get smaller gravitational deflection angle in the weak deflection limits (where $\frac{M}{b}$, $\frac{\lambda_{\text{DM}}}{b}$ are not large). Particularly, when the dark matter parameter is sufficiently large ($\lambda_{\text{DM}}>0.5M$), the gravitational deflection angle could become negative. These effects are mostly caused by the leading order contribution $-\frac{\lambda_{\text{DM}}}{b}-\frac{2\lambda_{\text{DM}}}{b}\cdot\ln\big(\frac{b}{|2\lambda_{\text{DM}}|}\big)$ in the analytical expression of the gravitational deflection angle. In addition, as dark matter parameters $\lambda_{\text{DM}}$ increases, the photon sphere radius and black hole shadow radius for charged RN black hole immersed in PFDM first diminish, and they begin to increase when dark matter parameters exceed the tuning points (which are $\lambda_{\text{DM}} \approx 0.5M$ for photon spheres and $\lambda_{\text{DM}} \approx 0.75M$ for black hole shadows). Furthermore, the Einstein ring angle of the Schwarzschild black hole immersed in PFDM reduces when dark matter parameter $\lambda_{\text{DM}}$ increases, and the Einstein ring angle enlarges when ratio
$ D_{\text{LS}}/D_{\text{OS}} $ becomes larger.

In addition, for the charged RN black hole immersed in PFDM, the black hole electric charge $Q$ has negative contributions to the gravitational deflection angle, photon sphere and black hole shadow. For a given fixed dark matter parameter $\lambda_{\text{DM}}$, black holes with smaller electric charge result in smaller gravitational deflection angles, unstable photon spheres and black hole shadows in the gravitational lensing. Furthermore, in the extreme charged RN black hole case ($Q=M$), the spacetime becomes horizonless and produces a naked singularity when dark matter parameter $0<\lambda_{\text{DM}}<M$, making the black hole shadow disappear in this region. 

In a summary, from the analytical and numerical results in the present study, the dark matter halo has non-negligible effects on the gravitational lensing of supermassive black holes. The gravitational deflection angle of light, photon sphere, black hole shadow and Einstein ring angle can be greatly changed in the presence of dark matter halos. However, the PFDM is still an over-simplified model to describe the detailed astronomical dark matter distributions in many galaxies. In the future, it is necessary to use the more precise and widely adopted phenomenological dark matter distributions (such as the NFW, Burkert, Brownstein, Moore models \cite{Navarro1996,Navarro1997,Burkert1995,Moore1998,Brownstein2006}) to further analyze the dark matter influences on the gravitational lensing of supermassive black holes in the center of galaxies.

\appendix

\acknowledgments

The authors thanks Song-Lin Lyu for helpful discussions on numerical calculations. This work is supported by the “zhitongche” program for doctors from Chongqing Science and Technology Committee (Grant No. CSTB2022BSXM-JCX0100), the Natural Science Foundation of Chongqing (Grant No. cstc2020jcyj-msxmX0879 and Grant No. CSTB2022NSCQ-MSX0932), the Scientific and Technological Research Program of Chongqing Municipal Education Commission (Grant No. KJQN202201126), and the Scientific Research Foundation of Chongqing University of Technology (Grants No. 2020ZDZ027).

\paragraph{Note added.} 
During the submission process, a very recent work pointed out the leading-order solution $u(\phi)=\frac{1}{r(\phi)} \approx \frac{\sin{\phi}}{b}$ we originally used in the integration of Gaussian curvature is unable to give all the correct terms in the second-order contributions of gravitational deflection angle \cite{YPHu2023}. The anonymous referee also gives the similar comments. We thank the anonymous referee and the authors of \cite{YPHu2023} for reminding.


\end{document}